\documentstyle[12pt,epsfig]{elsart}
\begin{document}

\newcommand{\re}{\mathop{\mathrm{Re}}}
\newcommand{\im}{\mathop{\mathrm{Im}}}
\newcommand{\D}{\mathop{\mathrm{d}}}
\newcommand{\I}{\mathop{\mathrm{i}}}

\noindent {\Large DESY 03--092 \hfill ISSN 0418-9833}

\noindent {\Large July 2003}

\vspace*{1cm}

\begin{frontmatter}

\journal{}
\date{}

\title{
Efficient frequency doubler for the soft X-ray SASE FEL at the TESLA Test
Facility}

\author[DESY]{J.~Feldhaus},
\author[DESY]{M.~K\"orfer},
\author[DESY]{T. M\"oller},
\author[DESY]{J. Pfl\"uger},
\author[DESY]{E.L.~Saldin},
\author[DESY]{E.A.~Schneidmiller},
and \author[Dubna]{M.V.~Yurkov}

\address[DESY]{Deutsches Elektronen-Synchrotron (DESY),
Notkestrasse 85, Hamburg, Germany}

\address[Dubna]{Joint Institute for Nuclear Research, Dubna,
141980 Moscow Region, Russia}

\begin{abstract}

This paper describes an effective frequency doubler scheme for SASE
free electron lasers. It consists of an undulator tuned to the first harmonic, a dispersion
section, and a tapered undulator tuned to the second harmonic.  The first
stage is a conventional soft X-ray SASE FEL. Its gain is controlled in
such a way that the maximum energy modulation of the electron beam at
the exit is about equal to the local energy spread, but still
far away from saturation.  When the electron bunch passes through the
dispersion section this energy modulation leads to effective
compression of the particles. Then the bunched electron beam enters the tapered
undulator and produces strong radiation in the process of coherent
deceleration.

We demonstrate that a frequency doubler scheme can be integrated
into the SASE FEL at the TESLA Test Facility at DESY, and will allow to
reach 3~nm wavelength with GW-level of output peak power. This would extend the 
 operating range of the FEL into the so-called water window  and significantly expand the
capabilities of the TTF FEL user facility.

\end{abstract}

\end{frontmatter}

\clearpage

\setcounter{page}{1}

\section{Introduction}

The soft X-ray FEL at the TESLA Test Facility at DESY (TTF FEL) will cover a spectral range between approximately 60 nm and 6 nm wavelength. The minimum wavelength of 6 nm is determined by the maximum electron beam energy of 1 GeV. It would be extremely interesting to extend this range into the so-called water window, i.e. the range between the K-Absorption edges of carbon and oxygen at 4.38nm and 2.34 nm, respectively. This would allow time-resolved studies of organic molecules above the carbon K-edge, and, for example, the investigation of thick, hydrated biological samples by X-ray microscopy without the need of shock-freezing
\cite{ttf-fel-cdr,ttf-fel-cdr-update}.  
There are several techniques under consideration to reach this goal:
A X-ray SASE FEL at shorter wavelengths, which requires higher beam
energy or a shorter undulator period, and the generation of harmonics
through a nonlinear mechanism driven by bunching at the fundamental.

There are basically two methods to up-convert the fundamental radiation frequency
via nonlinear harmonics. These methods are the generation of third
harmonic radiation in a planar SASE undulator through a nonlinear mechanism driven
by bunching at the fundamental frequency, and a two-undulator (second) harmonic
generation scheme, also referred to as the "after-burner" method.

SASE FELs are capable to produce powerful radiation not only at the
fundamental frequency, but also at higher harmonics. When a beam is
strongly bunched in the sinusoidal ponderomotive potential formed by
the undulator field and the radiation field of the fundamental
frequency, the electron beam density spectrum develops rich harmonic
contents.  Coherent radiation at the odd harmonics can be generated in
a planar undulator and significant power levels for the third harmonic
can be reached before the FEL saturates (see
\cite{sase-3rd-1,sase-3rd-2,sase-3rd-3,sase-3rd-4} and references
therein).  It is expected that the power of the transversely coherent
third-harmonic radiation can approach the 1\% level of the fundamental
power at the TTF. Since the nonlinear harmonic generation occurs
naturally in a planar undulator, no additional FEL hardware components
are required for this method.

An idea of using two undulators, with the second undulator resonant to
one of the harmonics of the first one, was considered in
\cite{bonifacio-3rd,ci,fa} (it is also referred to as the
``after-burner'' method). The first undulator is long enough to reach
saturation and produce strong spatial bunching in harmonics.  The
bunched beam generates coherent radiation in the second undulator which
follows immediately the first one.  The main problem with this approach
is the large induced energy spread which will be generated by the
bunching of the electron beam at the fundamental frequency. While this
energy spread is necessary for the bunching, it degrades the
performance of the radiator section at the harmonic frequency. Another
method to generate higher harmonics is the high-gain harmonic generation
scheme (see \cite{yu1,yu2} and references therein).  It should be noted that all
previous studies of such frequency multiplication schemes were performed
in the framework of the steady-state approach,  and the question arises
how effectively they work in the case of a SASE FEL.

\begin{figure}[tb]
\begin{center}
\epsfig{file=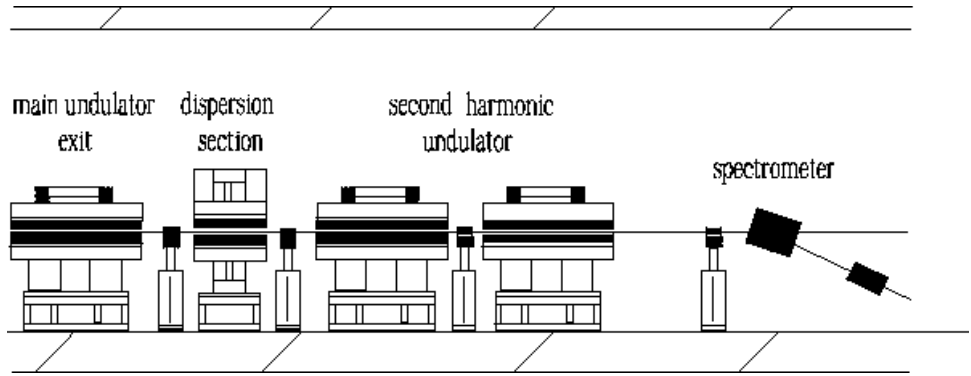,width=0.8\textwidth}
\end{center}
\caption{Side view of the electron beam transport system, showing the
location of the frequency doubler }
\label{fig:pp18c}
\end{figure}

In this paper we propose an effective frequency doubler scheme for a SASE
FEL. It consists of an undulator tuned to the first harmonic, a dispersion
section, and a tapered undulator tuned to the second harmonic (see
Fig.~\ref{fig:pp18c}).  The first stage is a conventional soft X-ray
SASE FEL.  The gain of the first stage is controlled in such a way that
the maximum energy modulation of the electron beam at the FEL exit is
about equal to the local energy spread, but still far away from
saturation.  When the electron bunch passes through the dispersion section this
energy modulation leads to effective compression of the particles. Then the
bunched electron beam enters a tapered undulator, and from the very
beginning produces strong radiation because of the large spatial
bunching. The strong radiation field produces a ponderomotive well which is
deep enough to trap the particles, since the original beam is relatively
cold.  The radiation produced by these captured particles increases the
depth of the ponderomotive well, and they are effectively decelerated.
As a result, much higher power can be achieved than for the case of a
uniform undulator. In addition, the output radiation exhibits excellent
spectral properties due to the suppression of the sideband growth in a
tapered undulator.

In this paper we analyze for the first time the frequency
multiplication scheme for a SASE FEL. Simulations using the
time-dependent code FAST \cite{fast}, upgraded for the simulation of higher
harmonics, provide a ``full physics'' description of the process. We
illustrate the operation of the proposed frequency doubler for the parameters
of the TTF FEL.  The result of our study is that a
frequency doubler can be implemented at the TESLA Test
Facility.  With an additional, 13.5~m long undulator (with 1.95~cm period
and 0.39~T peak magnetic field), the TTF FEL would be able to produce
radiation down to 3~nm wavelength with an output peak power in the GW range, and with
excellent spectral properties. The power of the third harmonic from this
device (i.e. at 1~nm wavelength) is still in the ten MW range, exceeding any other pulsed radiation source presently available at 1 nm, and sufficiently high for novel applications. In fact, this device 
approaches the operating range of the future TESLA XFEL.

The paper is organized as follows. Section 2 summarizes recent updates
of the TTF FEL parameters which refer mainly to a smaller local energy
spread than it was expected before.  Section 3 describes the operation of
the frequency doubler scheme. Section 4 discusses the 
integration of the frequency doubler scheme into the TESLA Test
Facility.  Appendixes 1 and 2 contain information on alternative ways
for attaining 3~nm wavelength at the TTF FEL using a uniform
undulator and an after-burner, respectively.

\section{6~nm operation of the SASE FEL at DESY}

\begin{figure}[b]
\begin{center}
\epsfig{file=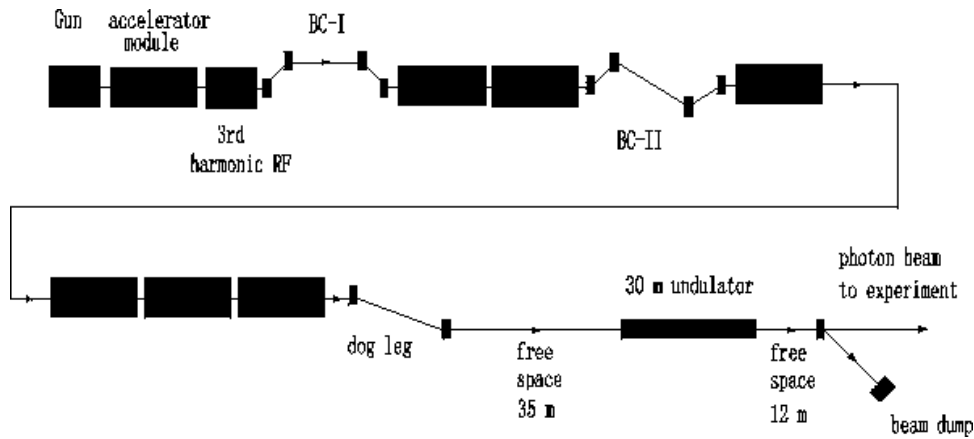,width=0.8\textwidth}
\end{center}
\caption{Schematic layout of the soft X-ray SASE FEL  at TTF, Phase 2
}
\label{fig:pp16}
\end{figure}

A soft X-ray SASE FEL will be commissioned as a user facility in 2004 at the
TESLA Test Facility at DESY, covering the wavelength range down to 6 nm.
The first description of the TESLA Test Facility FEL has been written
in 1995 \cite{ttf-fel-cdr}. An update of the TTF FEL design was fixed
in 2002 \cite{ttf-fel-cdr-update}. The main issues were the change of the undulator
focusing structure (separated instead of integrated)  and the bunch
compression system (removing the low-energy bunch compressor and
introducing a RF structure operating at the 3rd harmonic).

\begin{table}[b]
\caption{\sl
Parameters of electron beam at TTF
}
\medskip

\begin{tabular}{ l l }
\hline
Energy                      & 1000 MeV  \\
Peak current                & 2.5 kA     \\
Normalized rms emittance    & 2$\pi$ mm-mrad \\
rms energy spread           & 0.2 MeV    \\
rms bunch length            & 50 $\mu $m \\
External $\beta $-function  & 4.5 m      \\
rms beam size               & 68 $\mu $m \\
\hline
\end{tabular}

\label{tab:beampar}
\end{table}

\begin{figure}[b]

\epsfig{file=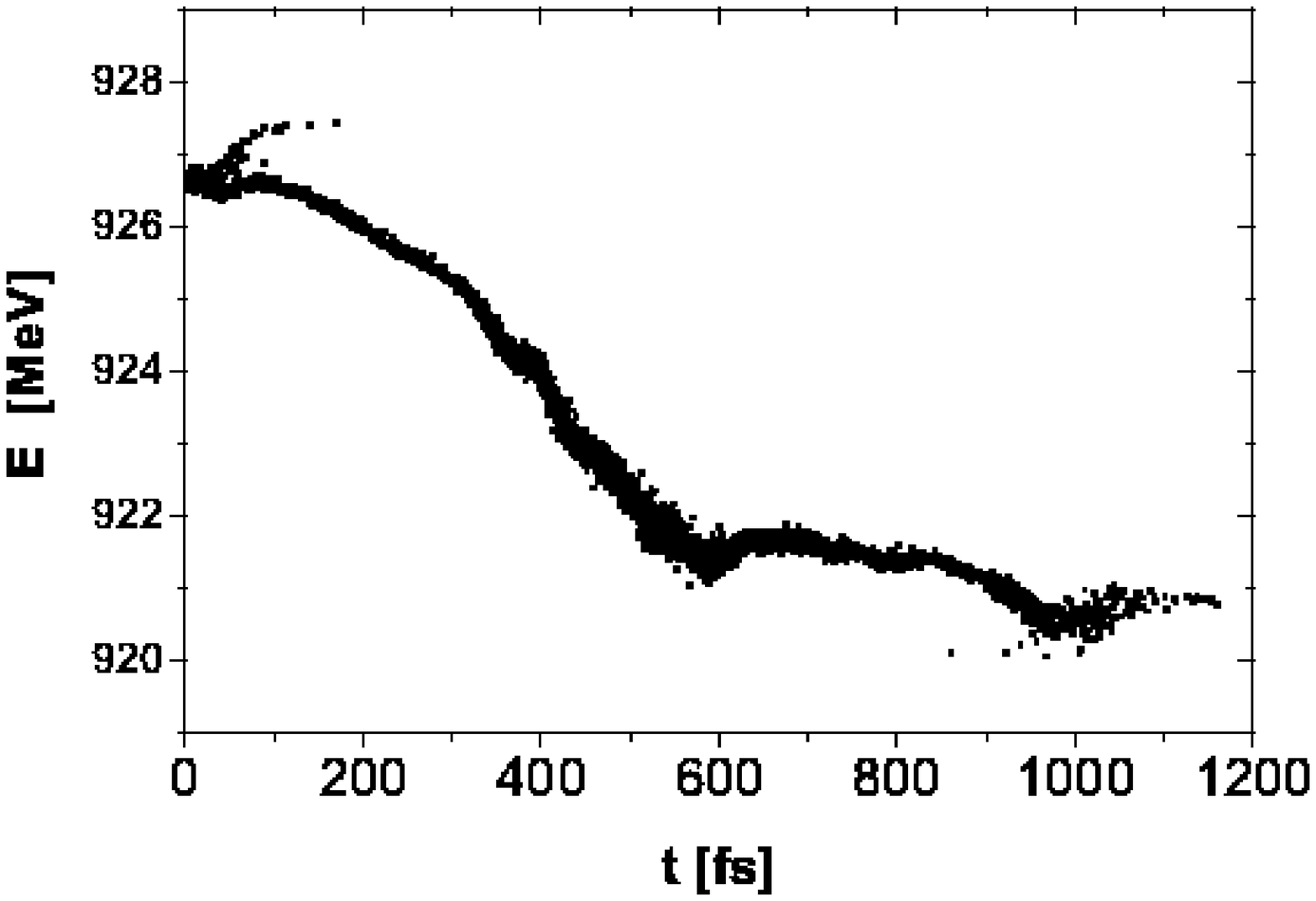,width=0.5\textwidth}

\vspace*{-62mm}

\hspace*{0.5\textwidth}
\epsfig{file=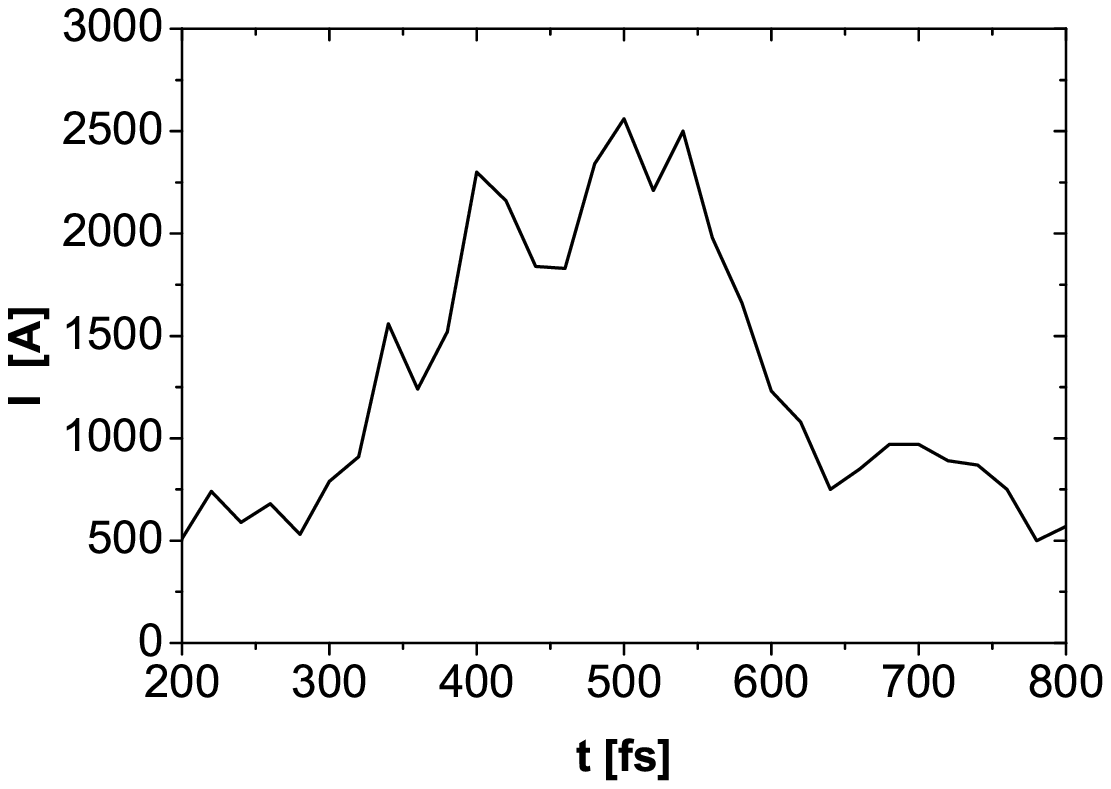,width=0.5\textwidth}

\epsfig{file=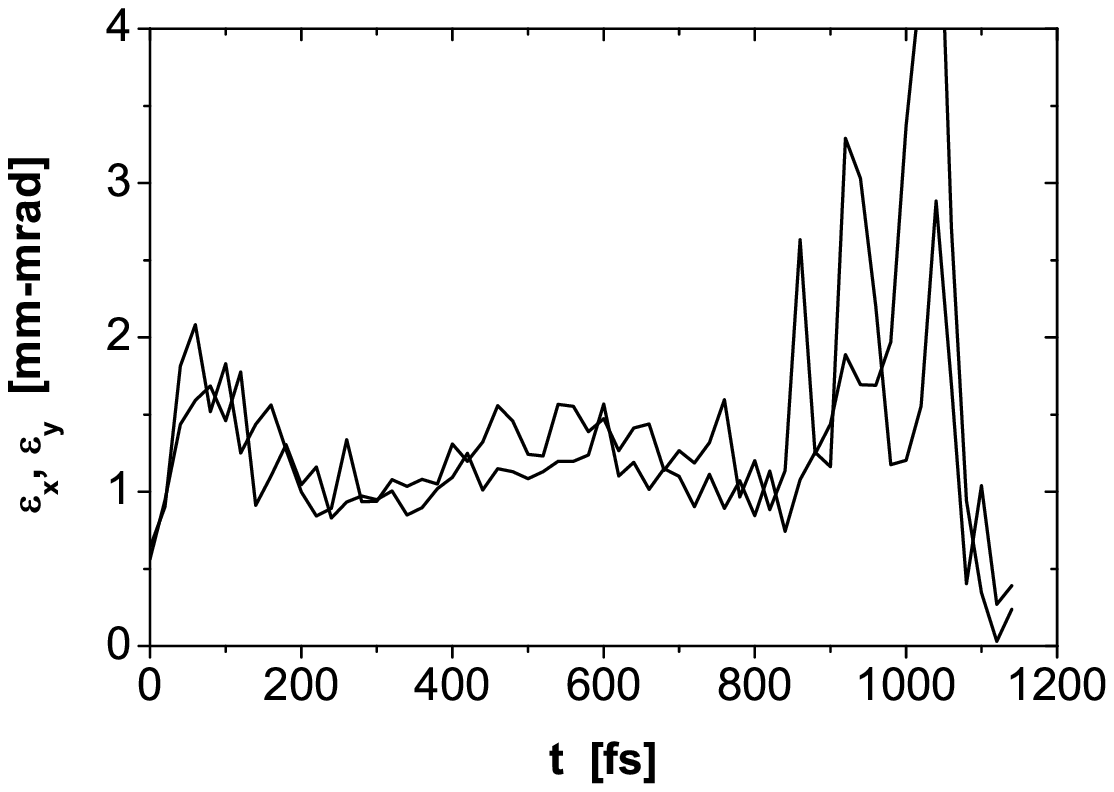,width=0.5\textwidth}

\vspace*{-62mm}

\hspace*{0.5\textwidth}
\epsfig{file=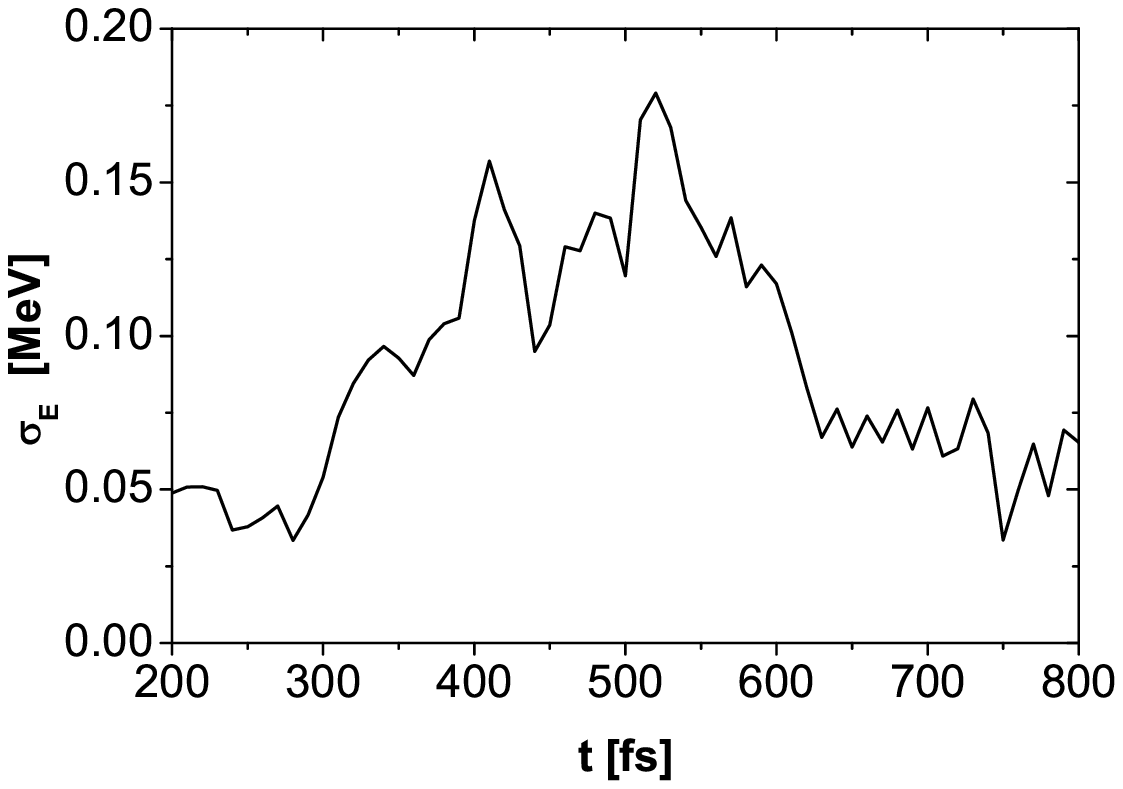,width=0.5\textwidth}

\caption{Nominal beam parameters for the TTF FEL, Phase 2: phase
space distribution, current along the bunch, slice emittance and slice
energy spread. The bunch charge is 1~nC \cite{limberg-epac02,piot-nsk}
}
\label{fig:piot-nsk}
\end{figure}

\begin{figure}[p]
\begin{center}
\epsfig{file=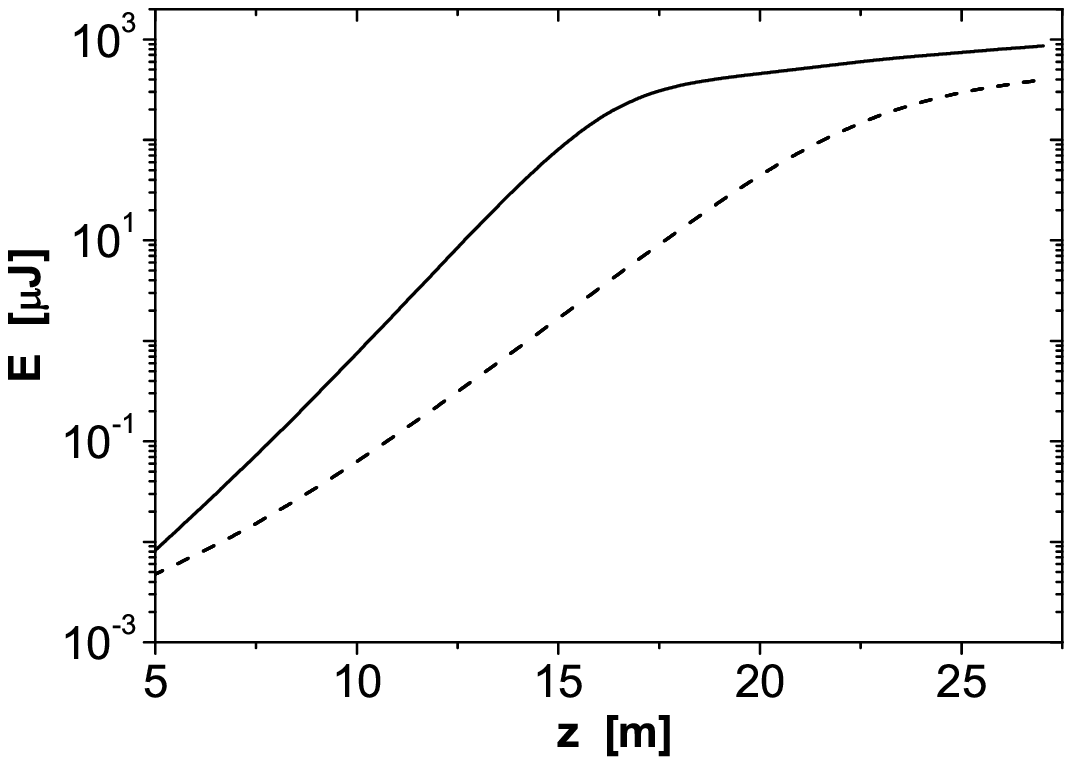,width=0.6\textwidth}
\end{center}
\caption{
Energy in the radiation pulse versus undulator length.
Nominal beam parameters for the TTF FEL, Phase 2 (see Tables
\ref{tab:beampar} and \ref{tab:6nm}). The radiation wavelength is equal to
6 nm.  The dashed line corresponds to the originally planned value of the local
energy spread of 1~MeV \cite{ttf-fel-cdr}
}

\label{fig:pz0102}


\epsfig{file=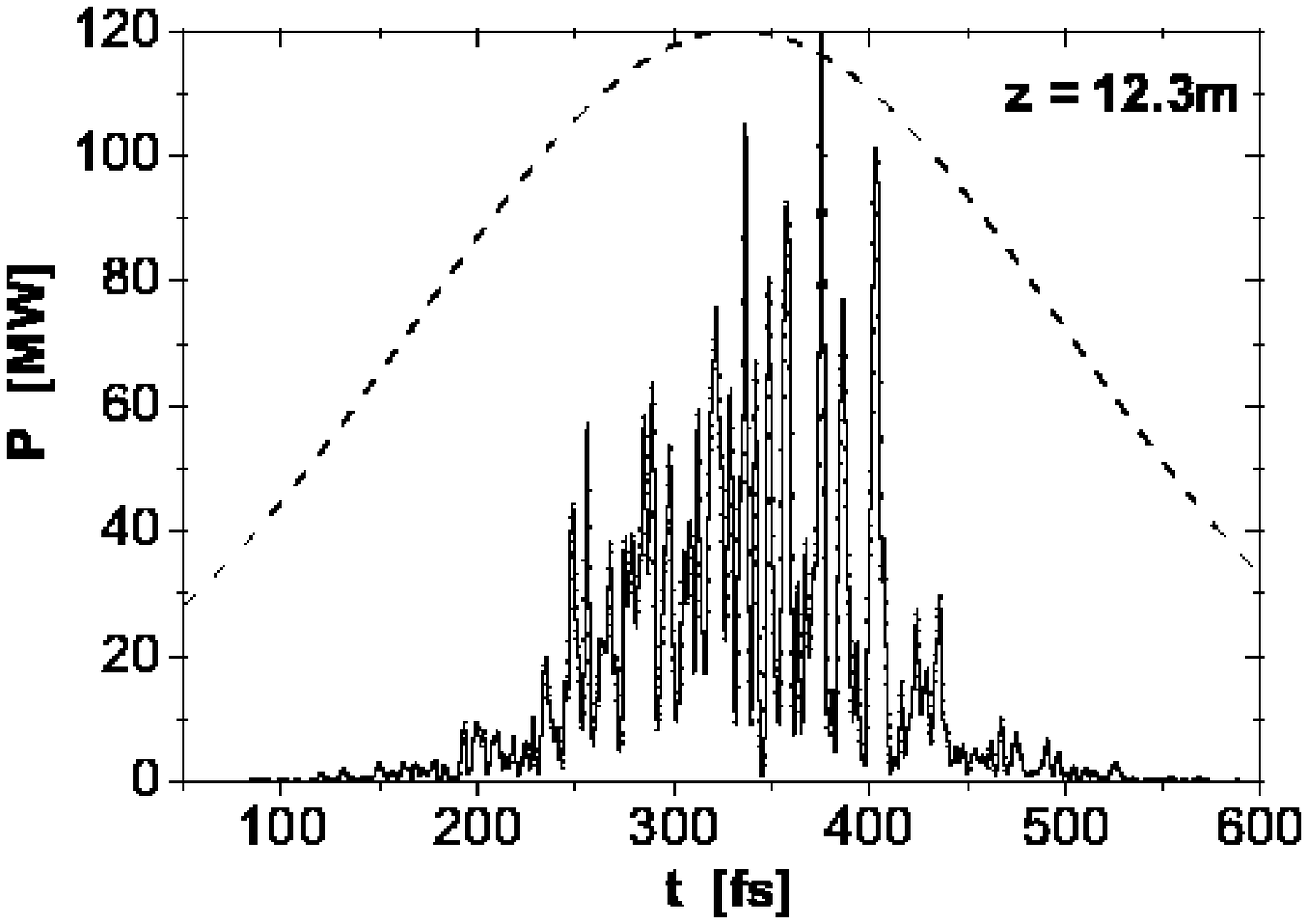,width=0.5\textwidth}

\vspace*{-62mm}

\hspace*{0.5\textwidth}
\epsfig{file=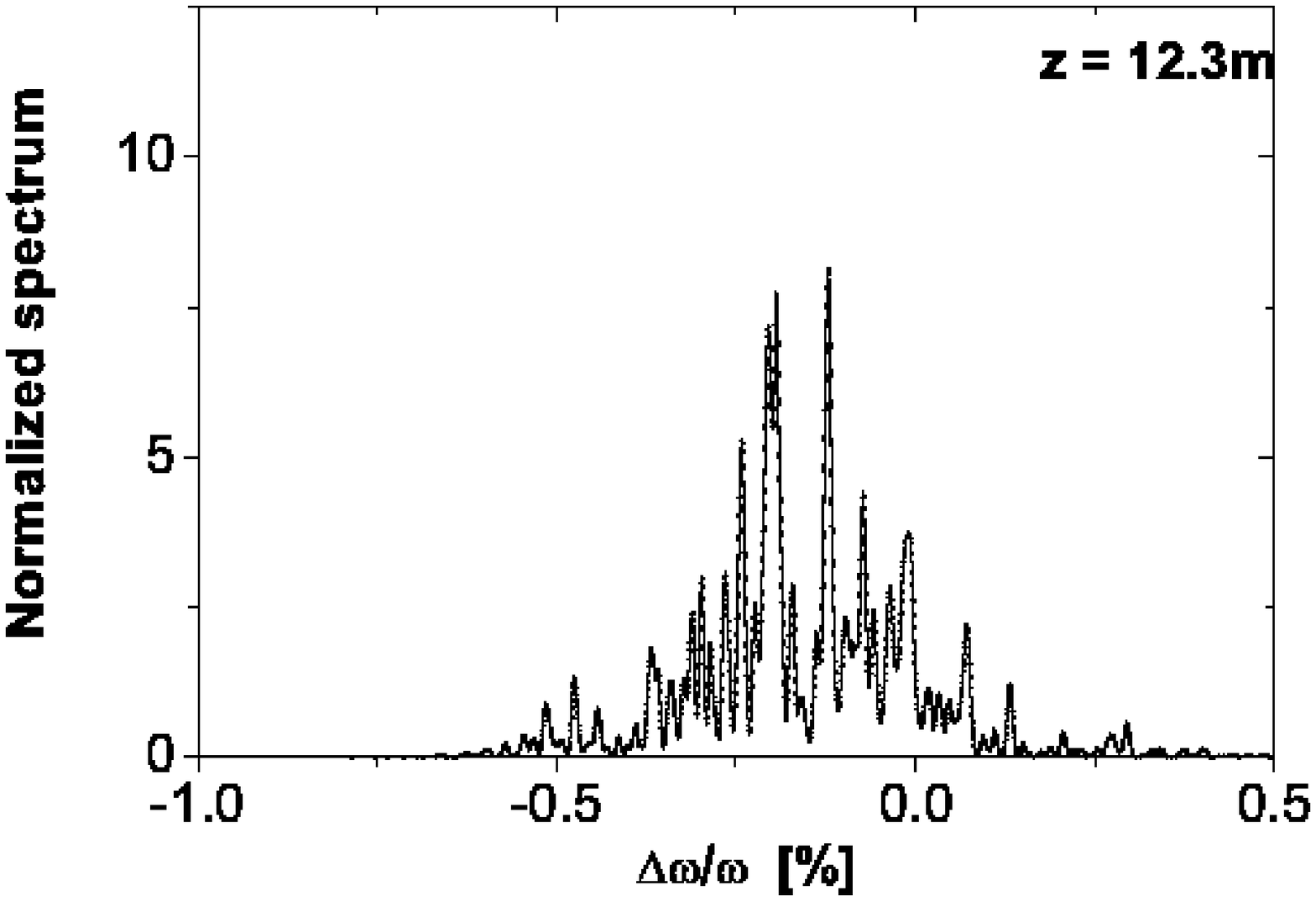,width=0.5\textwidth}

\epsfig{file=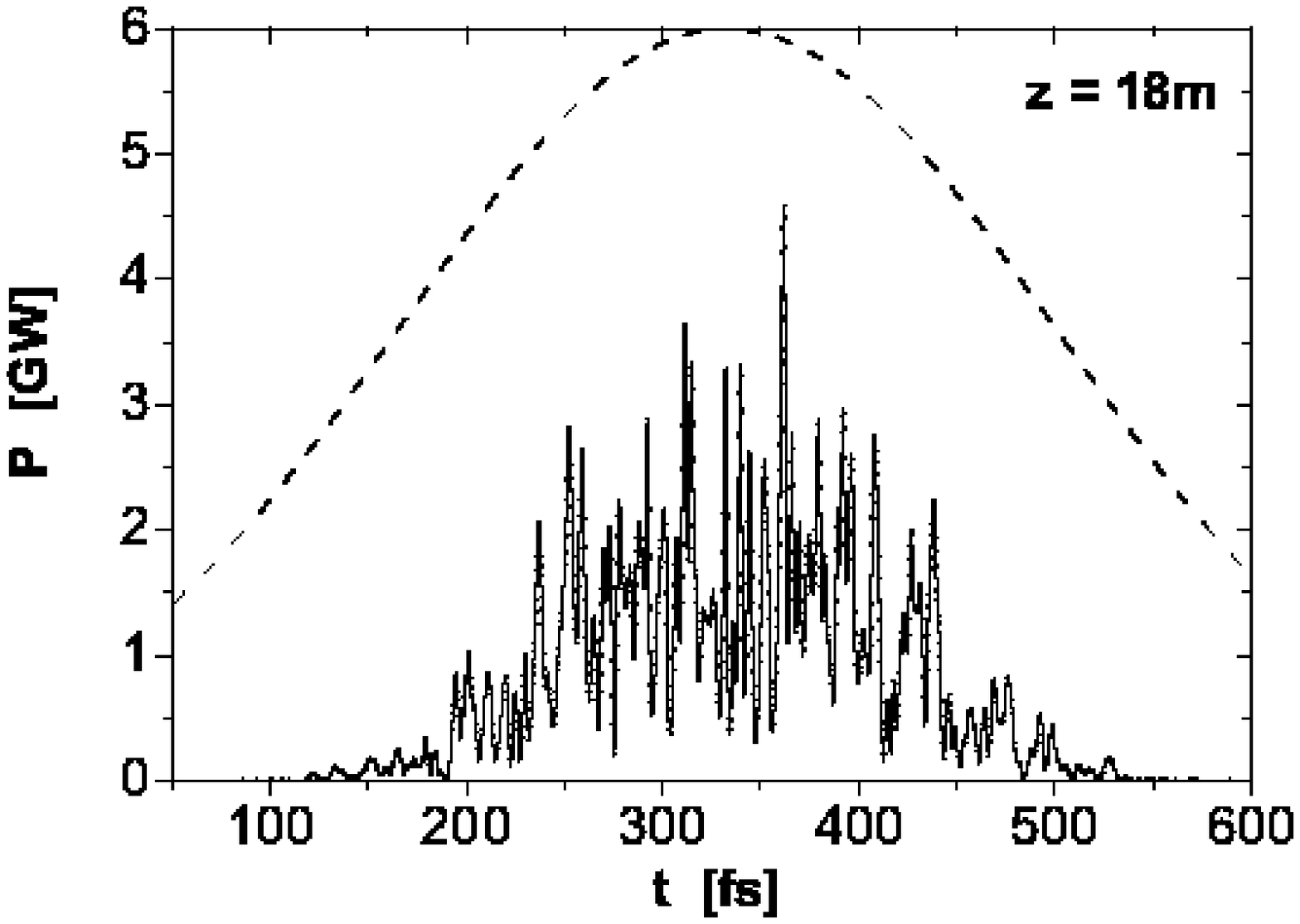,width=0.5\textwidth}

\vspace*{-62mm}

\hspace*{0.5\textwidth}
\epsfig{file=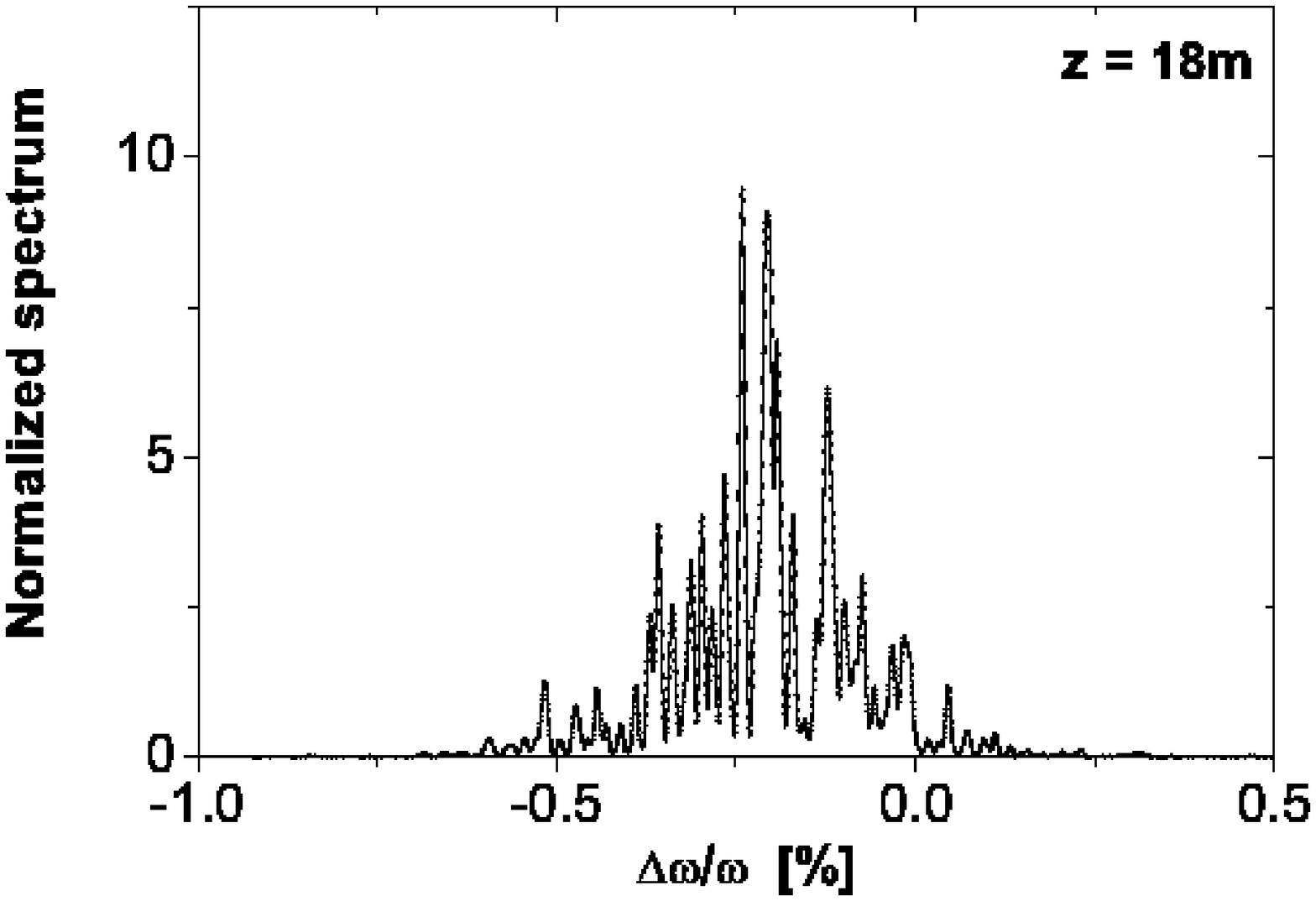,width=0.5\textwidth}

\caption{
Time structure (left column) and spectral structure (right column) of
the radiation pulse
for an undulator length 12.3~m (linear regime) and 18~m (saturation).
Nominal beam parameters for the TTF FEL, Phase 2 (see Tables
\ref{tab:beampar} and \ref{tab:6nm}).
The radiation wavelength is equal to 6~nm. The dashed line
shows the bunch profile
}
\label{fig:p0001037}
\end{figure}

\begin{table}[tb]
\caption{\sl
Nominal parameters of the TTF FEL, Phase 2 for 6~nm wavelength.
}
\medskip

\begin{tabular}{ l l }
\hline
\underline{Undulator} \\
\hspace*{10pt} Type                        & planar        \\
\hspace*{10pt} Period                      & 2.73 cm        \\
\hspace*{10pt} Gap                         & 12 mm         \\
\hspace*{10pt} Peak magnetic field         & 0.47 T        \\
\hspace*{10pt} Segment length              & 4.5 m         \\
\underline{Coherent radiation} \\
\hspace*{10pt} Wavelength                  & 6 nm       \\
\hspace*{10pt} Saturation length           & 18 m        \\
\hspace*{10pt} Peak power                  & 2 GW       \\
\hspace*{10pt} Bandwidth (FWHM)            & 0.4\%       \\
\hspace*{10pt} Pulse duration (FWHM)       & 200 fs       \\
\hline
\end{tabular}

\label{tab:6nm}
\end{table}

The analysis of recent experimental results obtained at the TTF FEL, Phase
1 \cite{ttf-sat-prl} led to the idea that the value of the local energy
spread is significantly less than it was expected originally. Only a
small value of the local energy spread allows one to obtain a high-peak
value of the bunch current after a single bunch compressor. Direct
measurements of the local energy spread gave an upper estimate below 5~keV at a beam current of about
100~A \cite{schlarb-pac03}. An extrapolation of this experimental result
to the project value of 2500~A beam current gives an estimate for
the local energy spread of about 100~keV. This is in agreement with
recent start-to-end simulations of the bunch in the TTF linac
\cite{limberg-epac02,piot-nsk} predicting the value of the local energy
spread within the lasing part of the bunch to be well below 200~keV
(see Fig.~\ref{fig:piot-nsk}).  Note that the original project value for
the local energy spread was 1~MeV
\cite{ttf-fel-cdr,ttf-fel-cdr-update}.  The local energy spread is one of
the essential parameters influencing the FEL operation, and we have to
analyze which impact it may have on the operation of the current version of
the TTF FEL and its future developments.

Tables~\ref{tab:beampar} and \ref{tab:6nm} show an updated list of
parameters for the SASE FEL at the TESLA Test Facility at DESY. The decrease of
the local energy spread results in shorter a saturation length as it is
seen in Fig.~\ref{fig:pz0102}. This is a consequence of the fact that the
previous design did not have sufficient safety margin with respect to
the energy spread. Figure~\ref{fig:p0001037} shows the time and spectral
structure of the radiation pulse for the linear and saturation regime.

With a small value of the local energy spread we fall in a different
region of physical parameters which reveals a possibility to implement
different FEL amplifier schemes. In particular, the application of
dispersion sections for beam bunching becomes very effective.

\section{Operation of a frequency doubler}

\begin{table}[b]
\caption{\sl
Parameters of the 
frequency doubler for the TESLA Test Facility FEL
}
\medskip

\begin{tabular}{ l l }
\hline
\underline{Undulator (1st harmonic)} \\
\hspace*{10pt} Type                        & planar        \\
\hspace*{10pt} Period                      & 2.73 cm        \\
\hspace*{10pt} Gap                         & 12 mm         \\
\hspace*{10pt} Peak magnetic field         & 0.47 T        \\
\hspace*{10pt} Segment length              & 4.5 m         \\
\hspace*{10pt} Undulator length            & 13.5 m        \\
\underline{Dispersion section} \\
\hspace*{10pt} Net compaction factor       & 1.5~$\mu $m \\
\underline{Undulator (2nd harmonic)} \\
\hspace*{10pt} Type                        & planar, tapered \\
\hspace*{10pt} Period                      & 1.95 cm        \\
\hspace*{10pt} Gap                         & 10 mm         \\
\hspace*{10pt} Peak magnetic field         & 0.39 T        \\
\hspace*{10pt} Segment length              & 4.5 m         \\
\hspace*{10pt} Undulator length            & 13.5 m        \\
\hspace*{10pt} Undulator tapering          & -0.14\%/m      \\
\underline{Coherent radiation} \\
\hspace*{10pt} Wavelength                  & 3 nm       \\
\hspace*{10pt} Energy per pulse            & 180 $\mu $J \\
\hspace*{10pt} Peak power                  & 1.5 GW       \\
\hspace*{10pt} Bandwidth (FWHM)            & 0.2\%       \\
\hspace*{10pt} Pulse duration (FWHM)       & 130 fs       \\
\hline
\end{tabular}

\label{tab:doubler}
\end{table}

\begin{figure}[tb]

\epsfig{file=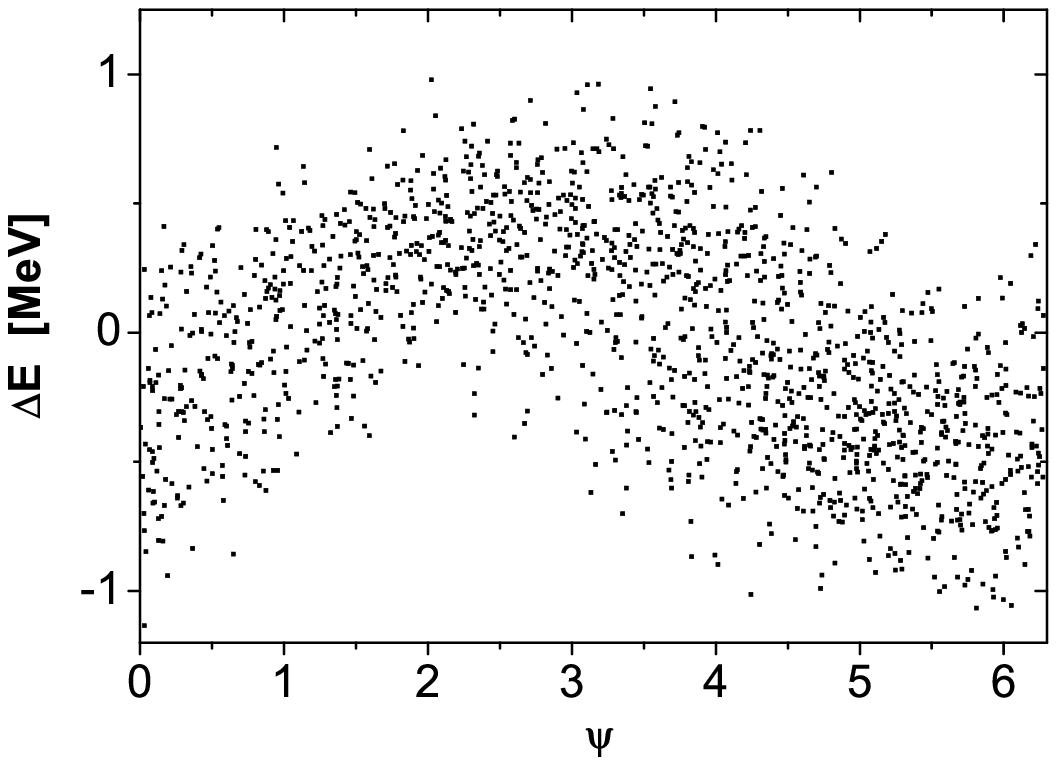,width=0.5\textwidth}

\vspace*{-62mm}

\hspace*{0.5\textwidth}
\epsfig{file=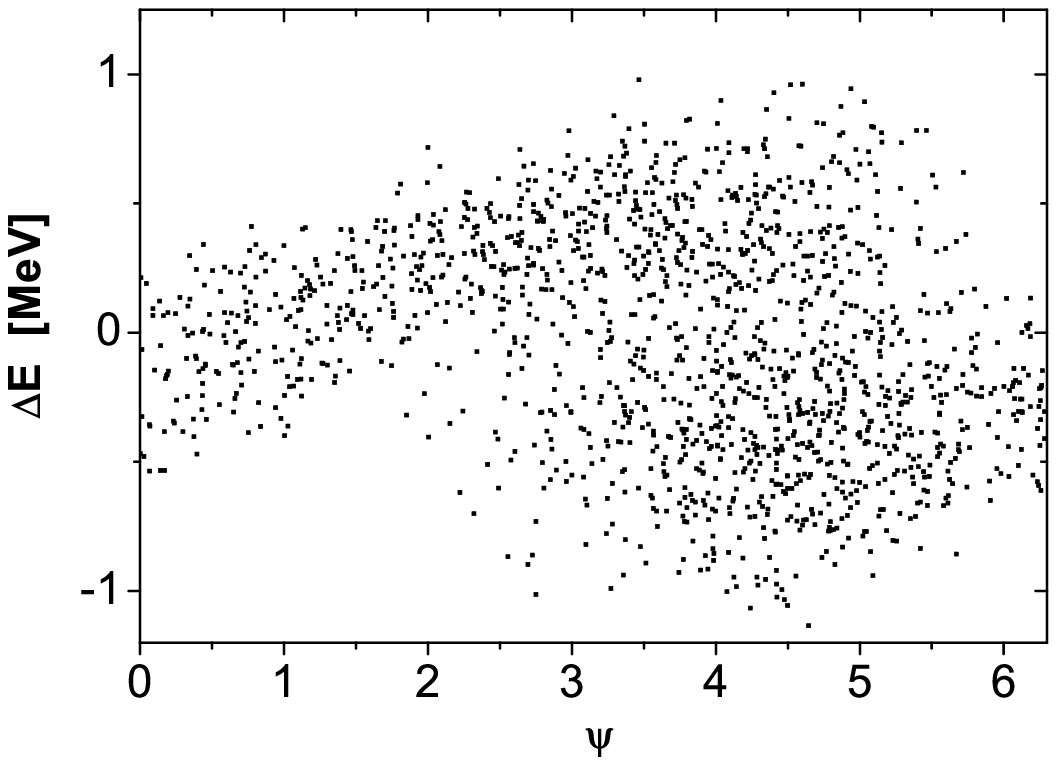,width=0.5\textwidth}

\epsfig{file=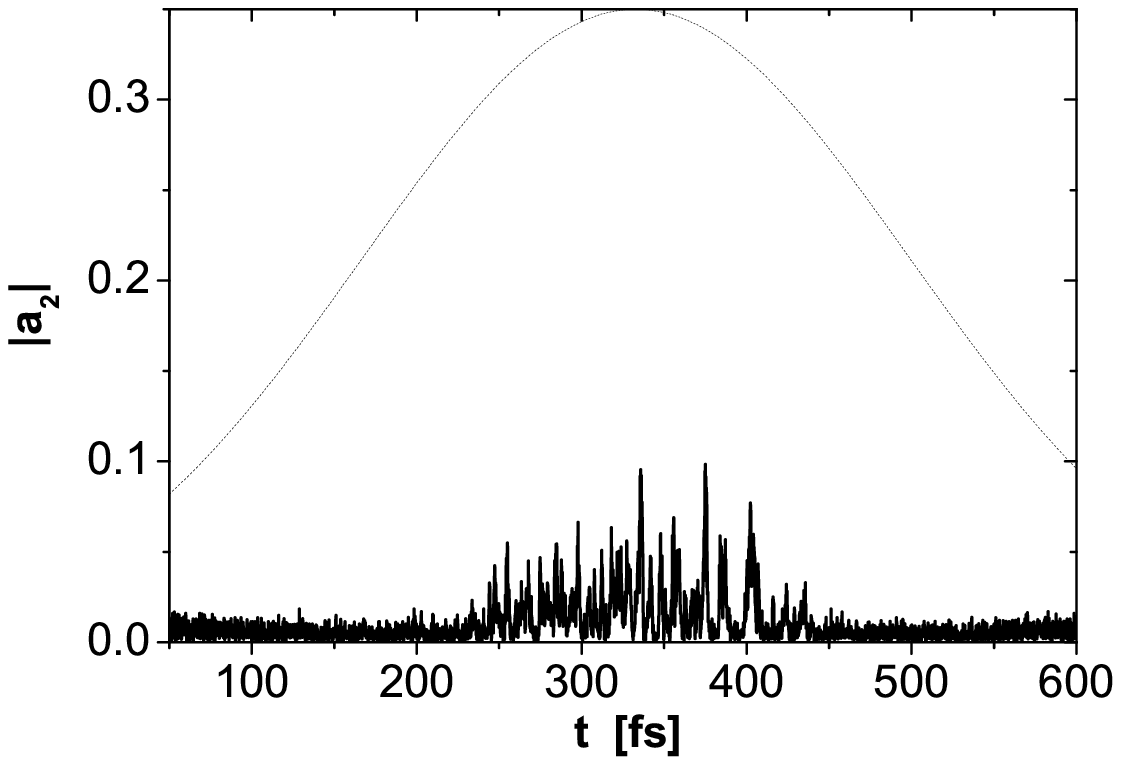,width=0.5\textwidth}

\vspace*{-62mm}

\hspace*{0.5\textwidth}
\epsfig{file=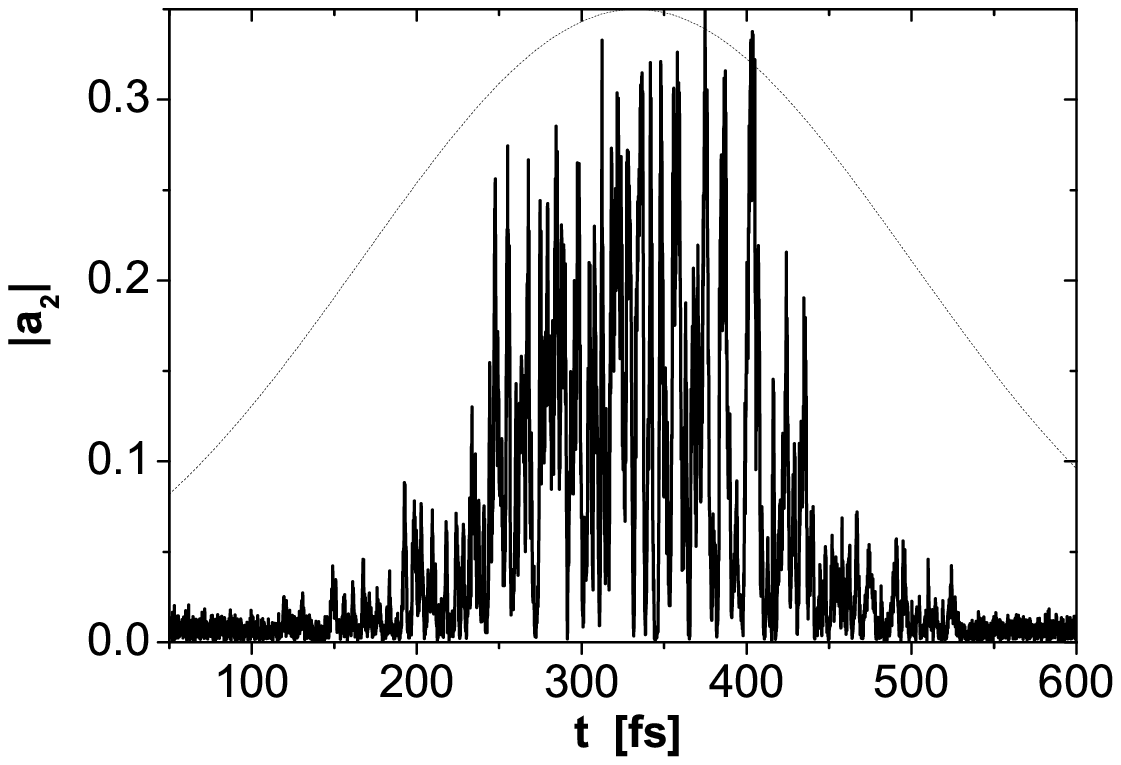,width=0.5\textwidth}

\caption{
Phase space distribution of the particles in a slice and second
harmonic of the beam density before (left column) and after (right
column) the dispersion section.  The radiation wavelength is equal to 6 nm.
The dashed line shows the bunch profile.
}
\label{fig:a25d15b}
\end{figure}

\begin{figure}[p]
\begin{center}
\epsfig{file=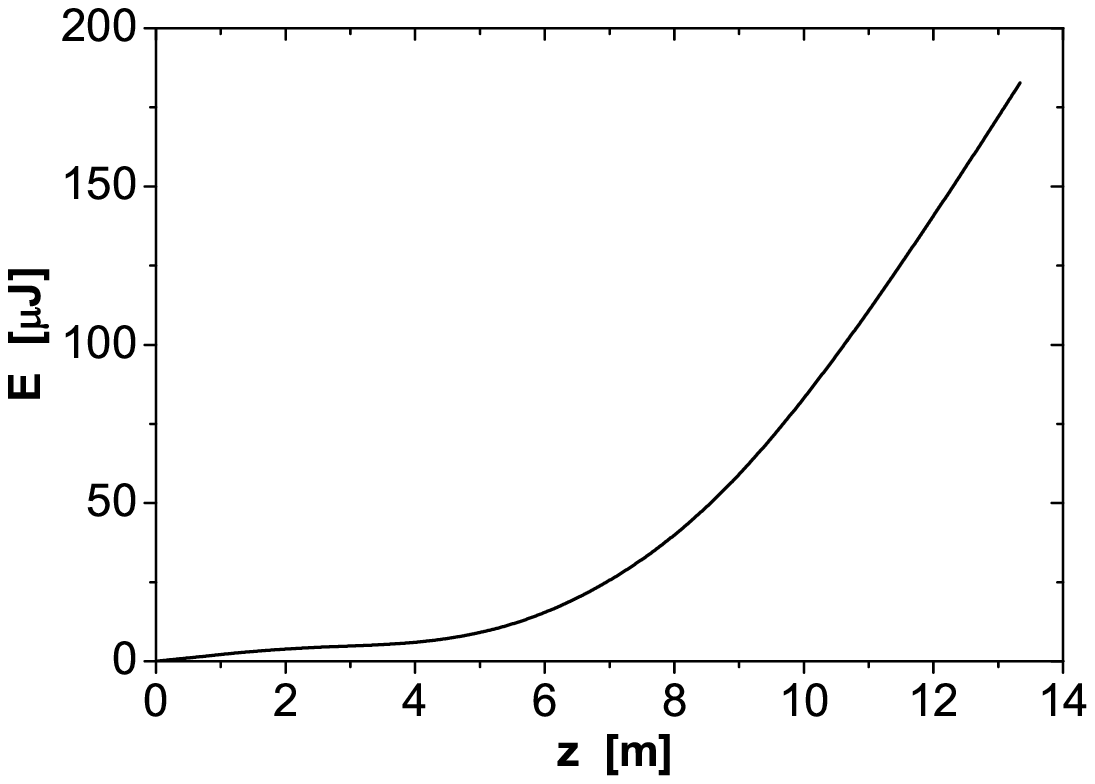,width=0.6\textwidth}
\end{center}
\caption{
Energy in the radiation pulse versus undulator length in the
frequency doubler
}
\label{fig:ez11}


\epsfig{file=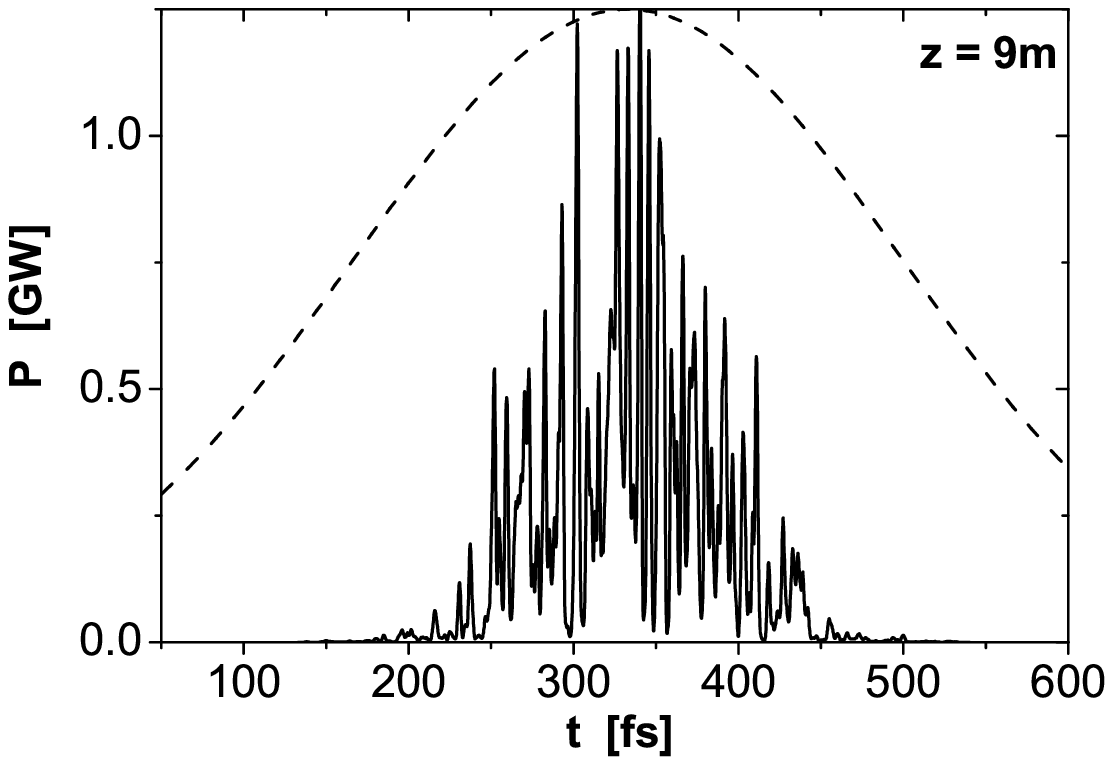,width=0.5\textwidth}

\vspace*{-62mm}

\hspace*{0.5\textwidth}
\epsfig{file=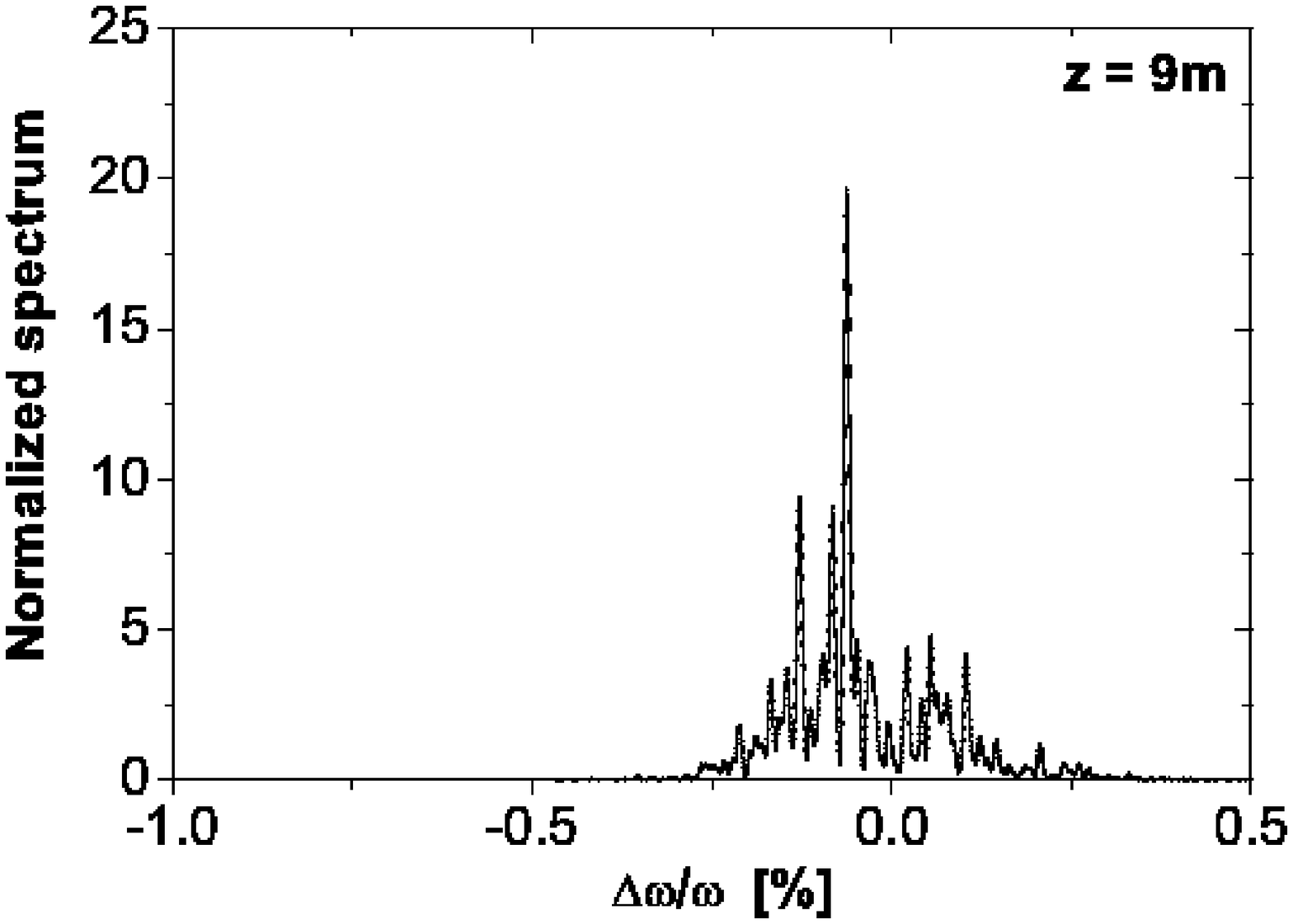,width=0.5\textwidth}

\vspace*{2mm}

\epsfig{file=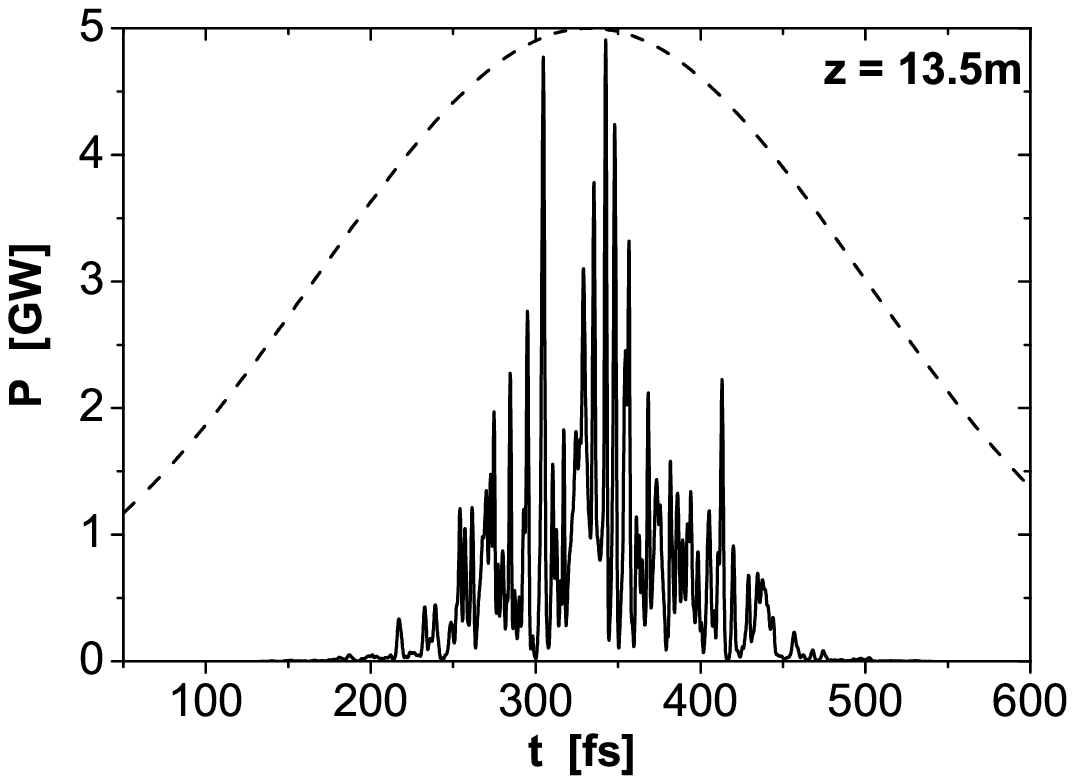,width=0.5\textwidth}

\vspace*{-62mm}

\hspace*{0.5\textwidth}
\epsfig{file=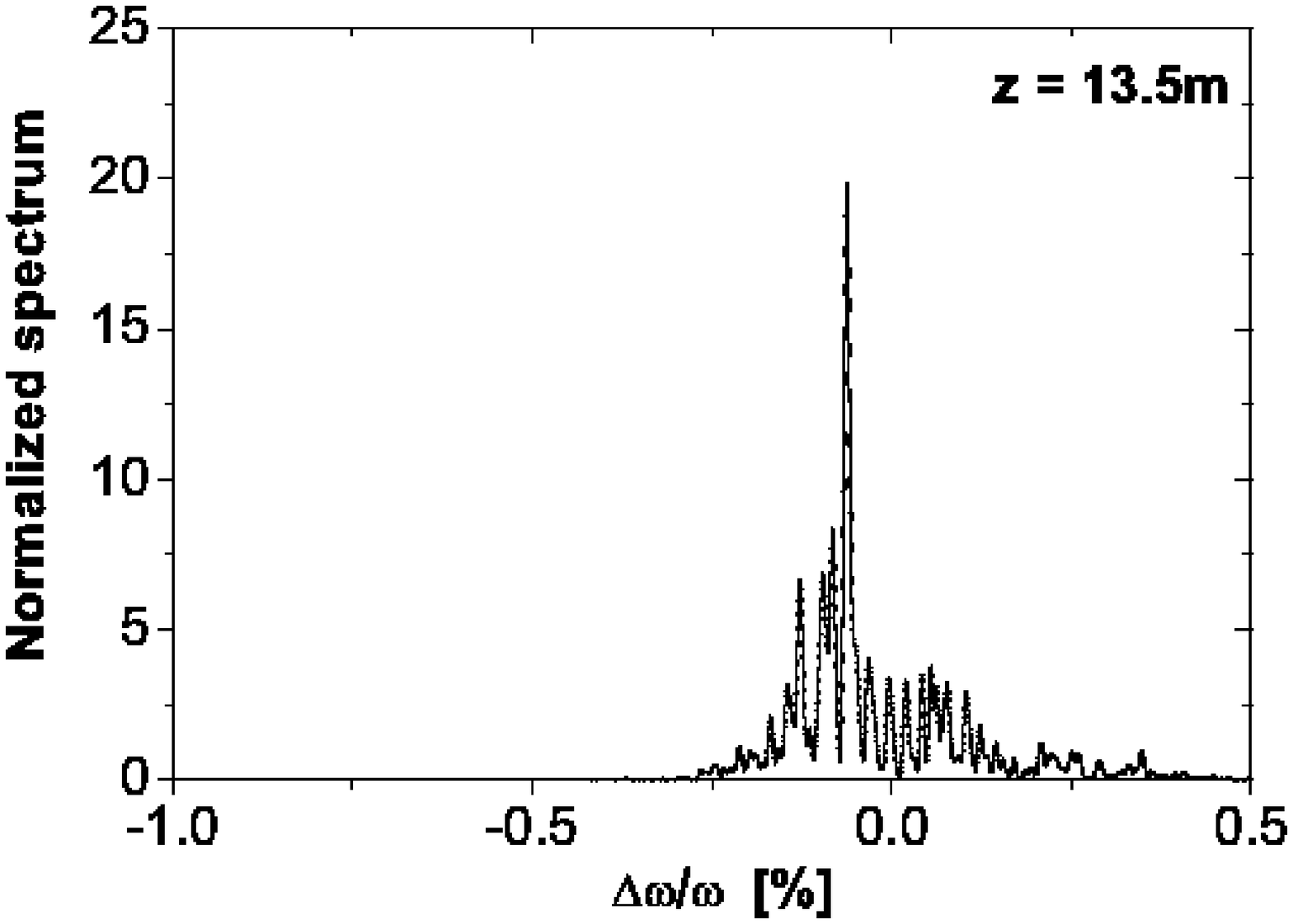,width=0.5\textwidth}

\caption{
Time structure (left column) and spectral structure (right column) of
the radiation pulse at a length of the frequency doubler of
9 and 13.5~meters.
The radiation wavelength is equal to 3 nm.
The dashed line shows the bunch profile.
}
\label{fig:p0011020}
\end{figure}

In this section we illustrate the operation of the frequency doubler for the
parameters of the TTF FEL, Phase 2. The beam parameters are given
in Table~\ref{tab:beampar}, and the parameters of the doubler undulator and the 
dispersion section are listed in Table~\ref{tab:doubler}. The resonance
wavelength in the main X-ray undulator is equal to 6~nm, and the doubler
undulator is tuned to the resonance wavelength of 3~nm.

The frequency doubler scheme operates as follows. The electron bunch enters the 
main X-ray undulator and produces SASE radiation at the wavelength of
6~nm.  During the amplification process the radiation power grows exponentially
with the undulator length. Simultaneously, the energy and density
modulation of the beam are growing. At the end of the X-ray undulator the beam
energy modulation is comparable with the local energy spread. In the
present example this takes place at an undulator length of 12.3~m.
The output characteristics of the radiation are illustrated with plots in
Fig.~\ref{fig:p0001037} (upper row). The upper left plot
in Fig.~\ref{fig:a25d15b} shows the phase space distribution of particles
for the slice corresponding to 330~fs time coordinate along the bunch.
Such a picture is typical for every spike. We see that the 
modulation amplitude at the
second harmonic is small (see lower left plot in this figure). On the
other hand, there is visible energy modulation with an amplitude of
about the value of the local energy spread. When the electron bunch passes
through the dispersion section this energy modulation leads to effective
compression of the particles as it is illustrated with plots in the
right column in Fig.~\ref{fig:a25d15b}. When the bunched beam enters the 
undulator tuned to the second harmonic, it immediately starts to
produce powerful radiation at the second harmonic. That is important,
but not the main feature of our proposal.
Let us study more closely the phase space distribution of the particles at
the exit of the dispersion section. We see that this distribution is
double-periodic with respect to the second harmonic. I.e., only each
second bucket is populated with particles. If we trace the evolution of the
FEL process in the uniform undulator, we find that ``thermalization''
takes place. The population of particles in the originally filled buckets is
reduced, and some of these particles travel into the ponderomotive well
of originally empty buckets. Simultaneously, the energy spread in the bunch
grows due to the FEL process. Finally, saturation occurs, and the properties of
the radiation are close to those described in Appendix~1.

Thus, another key idea of our proposal is to preserve the original bunching
of the beam and to organize an effective extraction of the energy from the
electron beam. This idea is simply realized by using a tapered undulator.
The process of amplification proceeds as follows. The bunched
beam (double-periodic) produces strong
radiation from the very beginning because of the large spatial bunching. The strong radiation
field produces a ponderomotive well which is deep enough to trap
particles, since the original beam is relatively cold.  The radiation
produced by these captured particles increases the depth of the
ponderomotive well, and they are effectively decelerated. The undulator
tapering preserves the synchronism of trapped particles and radiation,
and a significant fraction of the energy can be extracted. Simulations
using the time-dependent FEL code FAST \cite{fast} confirm this qualitative
picture.  Figure~\ref{fig:ez11} shows the evolution of the energy in the
radiation pulse in the tapered undulator. Despite of the original spiking
seeding the process of the second harmonic (see Fig.~\ref{fig:a25d15b}), we
effectively trap a significant fraction of the particles, and can achieve
much higher power than for the case of an untapered undulator (see
Appendix~1).  Figure~\ref{fig:p0011020} shows the temporal and spectral
structure of the radiation pulse after 9 and 13.5~meter of
second-harmonic undulator. One can see that the radiation pulse length is
about a factor of two shorter than that of a traditional SASE FEL (see
Fig.~\ref{fig:p0099045}). This is a consequence of the nonlinear
transformation provided by the dispersion section.
Another important feature of the radiation
from a tapered undulator is the significant suppression of the sideband
growth in the nonlinear regime (compare Figs.~\ref{fig:p0011020} and
\ref{fig:p0099045}). This means that in the proposed scheme the spectral brightness
of the radiation is increased proportionally to the radiation power.
In the case of a uniform undulator the peak brightness is reached at the
saturation point and is then reduced due to the sideband growth
\cite{book}.

\clearpage

\section{Implementation of a frequency doubler scheme at the TESLA TEST
Facility at DESY}

The TTF FEL is a pioneer user facility in the VUV and soft X-ray wavelength
range. The conceptual design was first elaborated in 1995
\cite{ttf-fel-cdr}. At that time only a single-pass SASE FEL
was considered. In 1998 it was decided to upgrade the
TTF FEL by a seeding option
\cite{seeding-option,seeding-option-cdr}. Other possible extensions
of the TTF FEL user facility making use of some space left behind the main FEL undulator have been discussed \cite{pump-pr1}.  

This free space downstream of the main
undulator (see Fig~\ref{fig:pp16}) would be sufficient for a 9~meter long
($2\times 4.5$~m sections) second harmonic radiator. It is planned now
that the main X-ray undulator will consist of six sections.  This is mainly
dictated by the wish to have sufficient safety margin during the first
commissioning of the SASE FEL. However, our analysis in the present paper
shows that for the expected electron beam parameters five undulator modules will be more
than sufficient. This means that in the future there will be space for
three modules of the frequency doubler. This will allow
to realize the full frequency doubler and to extend the 
wavelength range of the TTF FEL down to 3~nm with GW-level power without changing the
present TTF layout.  Additional hardware to be manufactured is a
dispersion section and a 13.5~m long undulator.  Since the required strength
of the dispersion section is small, its design may be similar to that
of a phase shifter for the X-ray SASE FEL \cite{phase-shifter}. It will also be
necessary to include a linear tapering of 0.14\%/m ( in the 13.5~m long undulator with 1.95~cm period and  0.39~T peak magnetic field.  This does not appear to be a serious
problem.  In conclusion we can state that these are minor expenses in
view of the significant extension of the capabilities of the TTF FEL user
facility.

\section{Conclusion}

In this paper we have described an effective frequency doubler for SASE FELs.  For the
first time the frequency multiplication scheme has been
analyzed for SASE FELs. To be specific, we have illustrated the
proposed FEL scheme using the parameters of the TESLA Test Facility. It is shown that
a frequency doubler allows to reach the water window
and to produce GW-level radiation pulses.  It is important to
note that a frequency doubler would be well compatible with the
seeding option, and would transfer all the advantages of the seeding
option to twice the photon energy
\cite{seeding-option,seeding-option-cdr}. In general, the frequency doubler
scheme has several significant advantages over
traditional SASE FELs with uniform undulators:

\begin{itemize}

\item shorter total magnetic length;

\item a possibility to attain higher output power and brightness of the
radiation;

\item shorter radiation pulse duration.

\end{itemize}

The realization of a frequency doubler at the TTF is considered as
extremely important not only for reaching the water window. It may
also be relevant for realizing the X-ray FEL user facility in the 0.1 nm range.  The use of
frequency doublers at a large-scale facility could result in a significant
reduction of the project costs. In addition, the safety margin for facility
operation becomes more relaxed.

\section*{Acknowledgments}

We thank J.~Krzywinski, J.~Rossbach, and M.~Tischer for many useful
discussions.  We thank C.~Pagani, J.R.~Schneider, and D.~Trines for their
interest in this work.

\clearpage

\section*{Appendix 1: 3~nm option with uniform undulator}

\begin{figure}[p]
\begin{center}
\epsfig{file=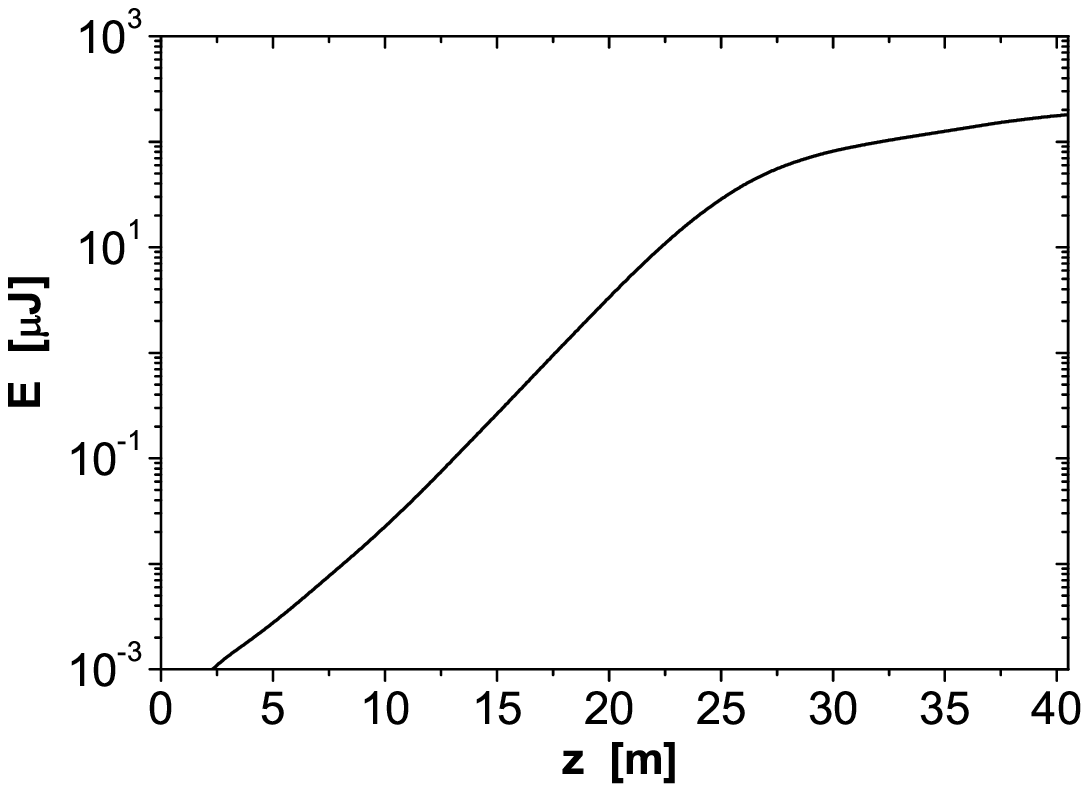,width=0.6\textwidth}
\end{center}
\caption{
Energy in the radiation pulse versus undulator length for a
3~nm option with uniform undulator
}
\label{fig:ez99}


\epsfig{file=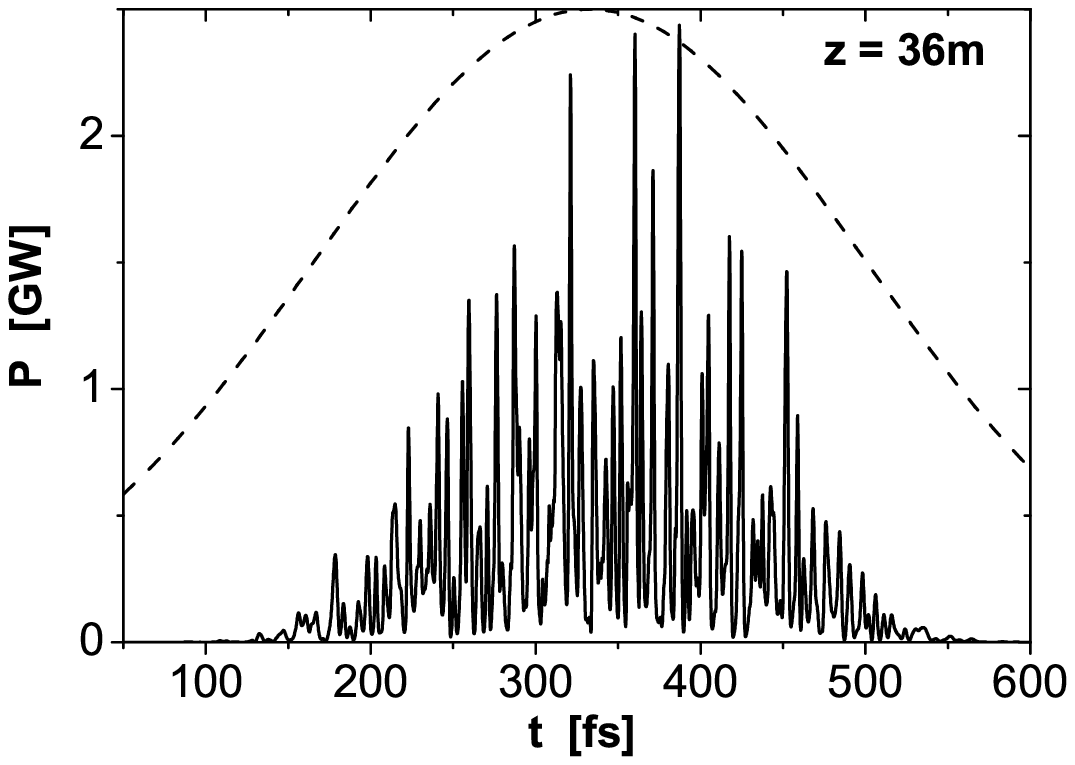,width=0.5\textwidth}

\vspace*{-62mm}

\hspace*{0.5\textwidth}
\epsfig{file=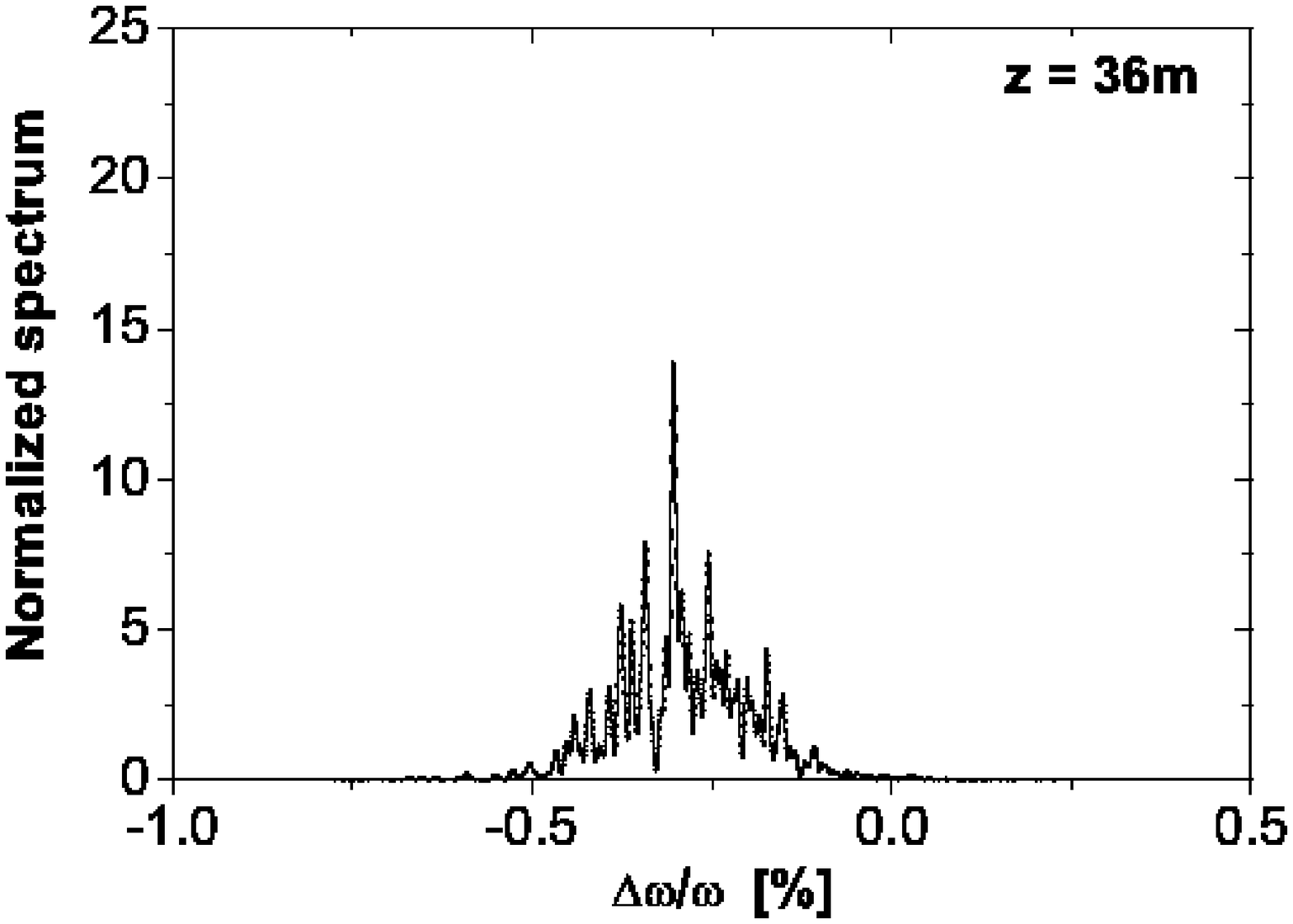,width=0.5\textwidth}

\vspace*{2mm}

\epsfig{file=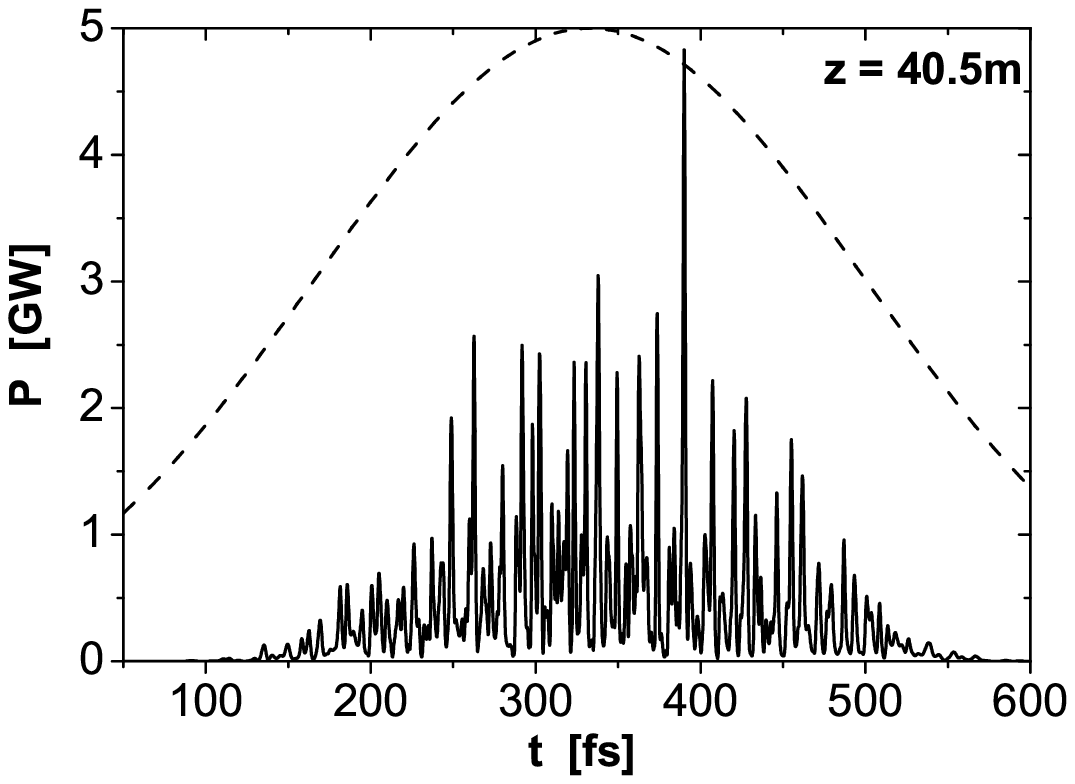,width=0.5\textwidth}

\vspace*{-62mm}

\hspace*{0.5\textwidth}
\epsfig{file=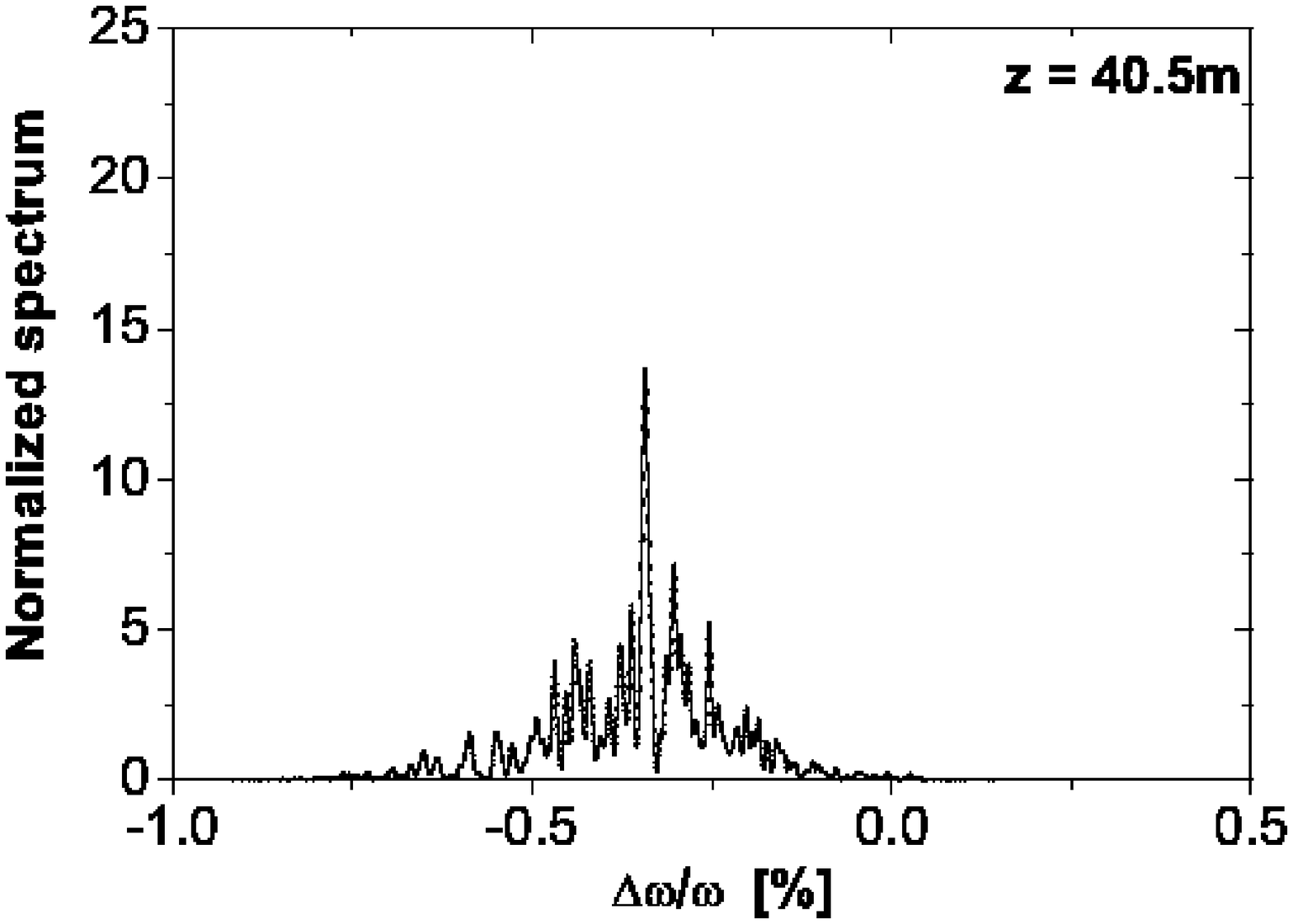,width=0.5\textwidth}

\caption{
Time structure (left column) and spectral structure (right column) of
the 3~nm radiation pulse for a uniform undulator at a distance of 36 and 40.5~meters.
The dashed line shows the bunch profile.
}
\label{fig:p0099045}
\end{figure}

The maximum energy of the TTF linac is limited to 1~GeV.
The present TTF undulator has a period of 2.73~cm and a peak magnetic field of 0.47~T
which limits the minimum wavelength to 6~nm. In
principle, one can consider a scenario of constructing a short-period
undulator allowing to generate wavelengths down to 3~nm. Such an option
was not studied previously because of the large value of 1~MeV for the expected local
energy spread. However, in view of the change of this parameter it
seems to be reasonable to perform such a study in order to have a
possibility for comparison of different approaches for attaining 3~nm
wavelength.  This section illustrates the option of a 3~nm SASE FEL with a
uniform undulator.  The parameters of the electron beam are presented in
Table~\ref{tab:beampar}, and those of the undulator are given in
Table~\ref{tab:3nm}.

Figure~\ref{fig:ez99} shows the evolution of the energy in the radiation
pulse along the undulator. The growth of the radiation energy starts to
saturate at an undulator length of about 36~m, and the radiation
energy is about 100~$\mu $J. Further increase of the undulator length
leads to a slow increase of the radiation energy, while the spectral width is
increased due to sideband growth in the nonlinear regime (see
Fig.~\ref{fig:p0099045}).

\begin{table}[h]
\caption{\sl
3~nm option with uniform undulator
}
\medskip

\begin{tabular}{ l l }
\hline
\underline{Undulator}\\
\hspace*{10pt} Type                        & planar \\
\hspace*{10pt} Period                      & 1.95 cm        \\
\hspace*{10pt} Gap                         & 10 mm         \\
\hspace*{10pt} Peak magnetic field         & 0.39 T        \\
\hspace*{10pt} Segment length              & 4.5 m         \\
\hspace*{10pt} Undulator length            & 36 m         \\
\underline{Coherent radiation} \\
\hspace*{10pt} Wavelength                  & 3 nm       \\
\hspace*{10pt} Energy per pulse            & 100 $\mu $J \\
\hspace*{10pt} Peak power                  & 500 MW       \\
\hspace*{10pt} Bandwidth (FWHM)            & 0.2\%       \\
\hspace*{10pt} Pulse duration (FWHM)       & 230 fs       \\
\hline
\end{tabular}

\label{tab:3nm}
\end{table}

\clearpage

\section*{Appendix 2: 3~nm option with after-burner}

\begin{figure}[p]

\epsfig{file=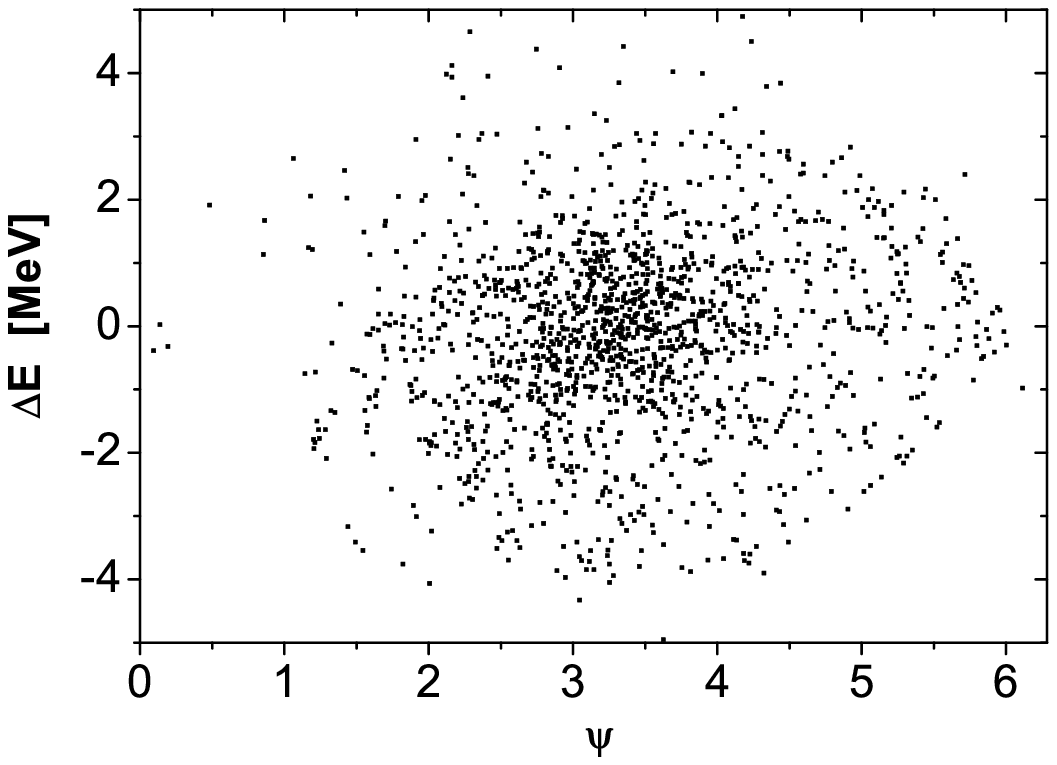,width=0.5\textwidth}

\vspace*{-62mm}

\hspace*{0.5\textwidth}
\epsfig{file=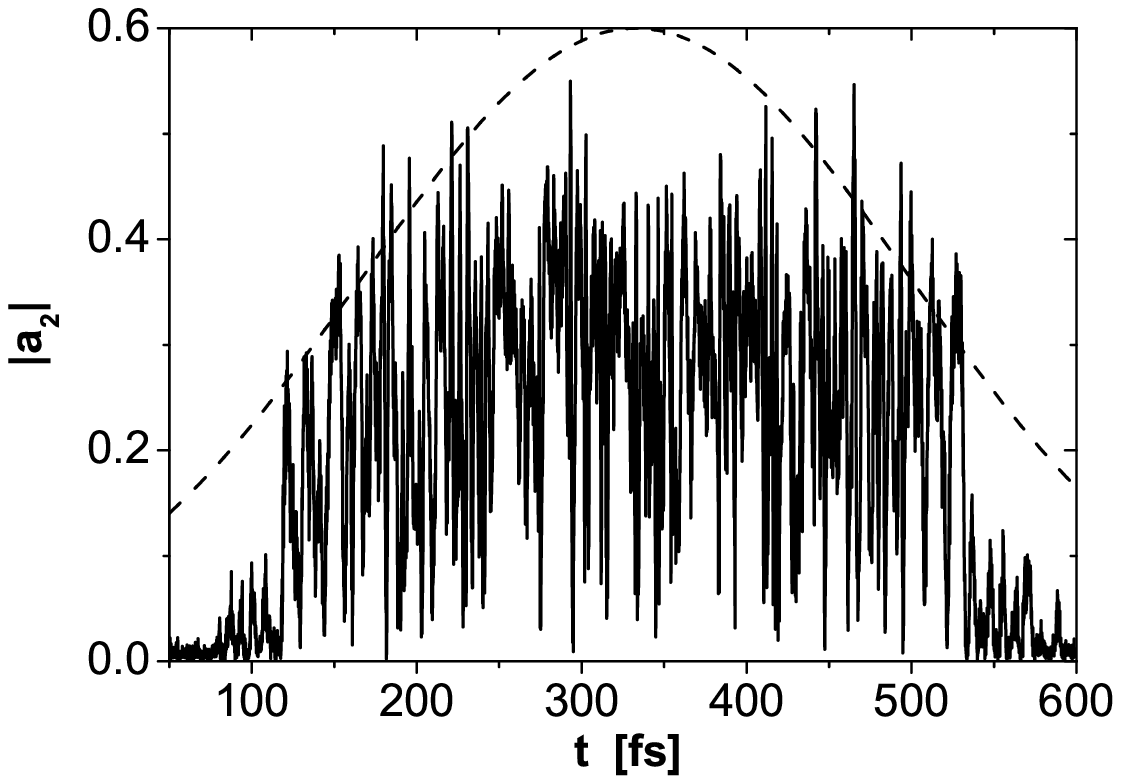,width=0.5\textwidth}

\caption{
Phase space distribution of the particles in a slice (left plot) and
amplitude of the second harmonic (right plot) at the exit of the
main undulator radiating at 6 nm wavelength.
The SASE FEL operates at saturation.
Nominal beam parameters for TTF FEL, Phase 2 (see Table
\ref{tab:6nm}) have been used for the simulation. The dashed line shows the bunch profile.
}
\label{fig:ab40a2}

\begin{center}
\epsfig{file=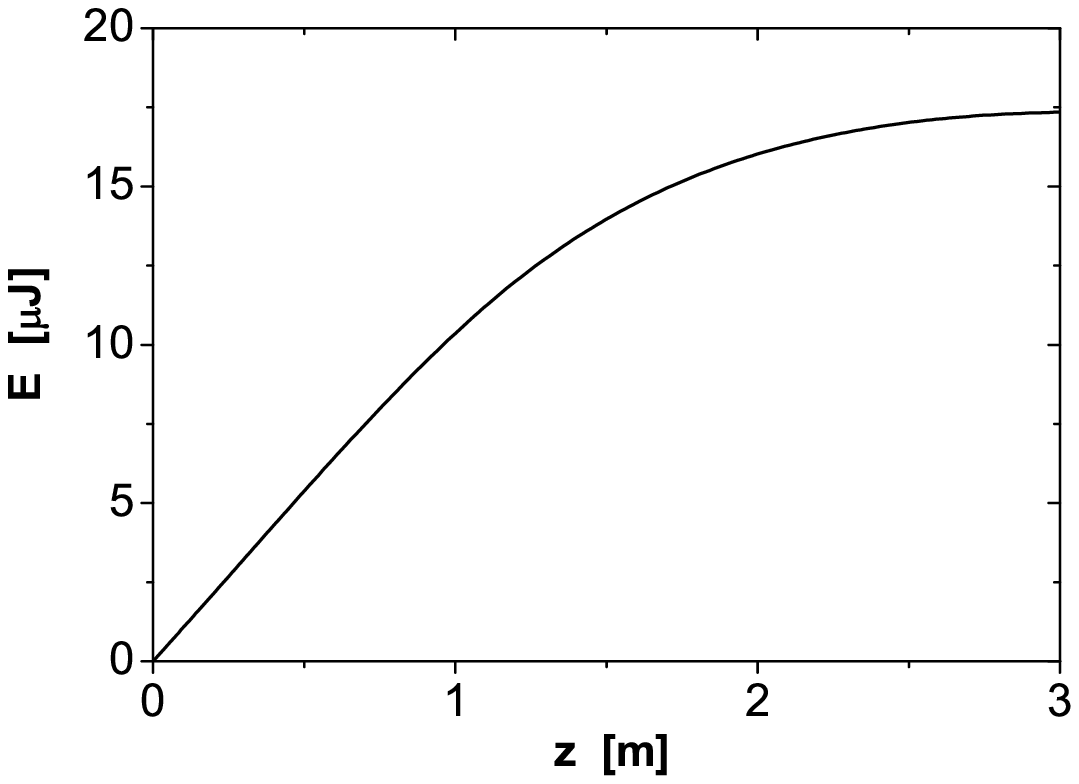,width=0.6\textwidth}
\end{center}
\caption{
Energy in the radiation pulse versus undulator length in the
after-burner.
The radiation wavelength is equal to 3 nm.
}
\label{fig:ez21}


\epsfig{file=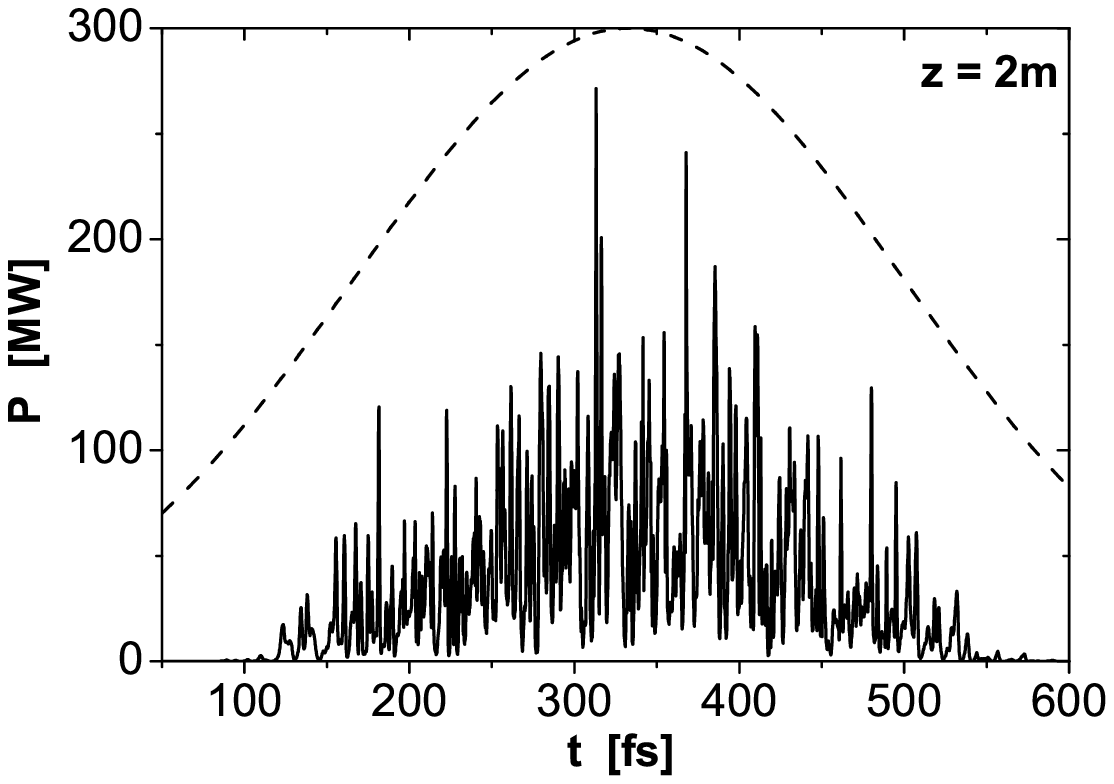,width=0.5\textwidth}

\vspace*{-62mm}

\hspace*{0.5\textwidth}
\epsfig{file=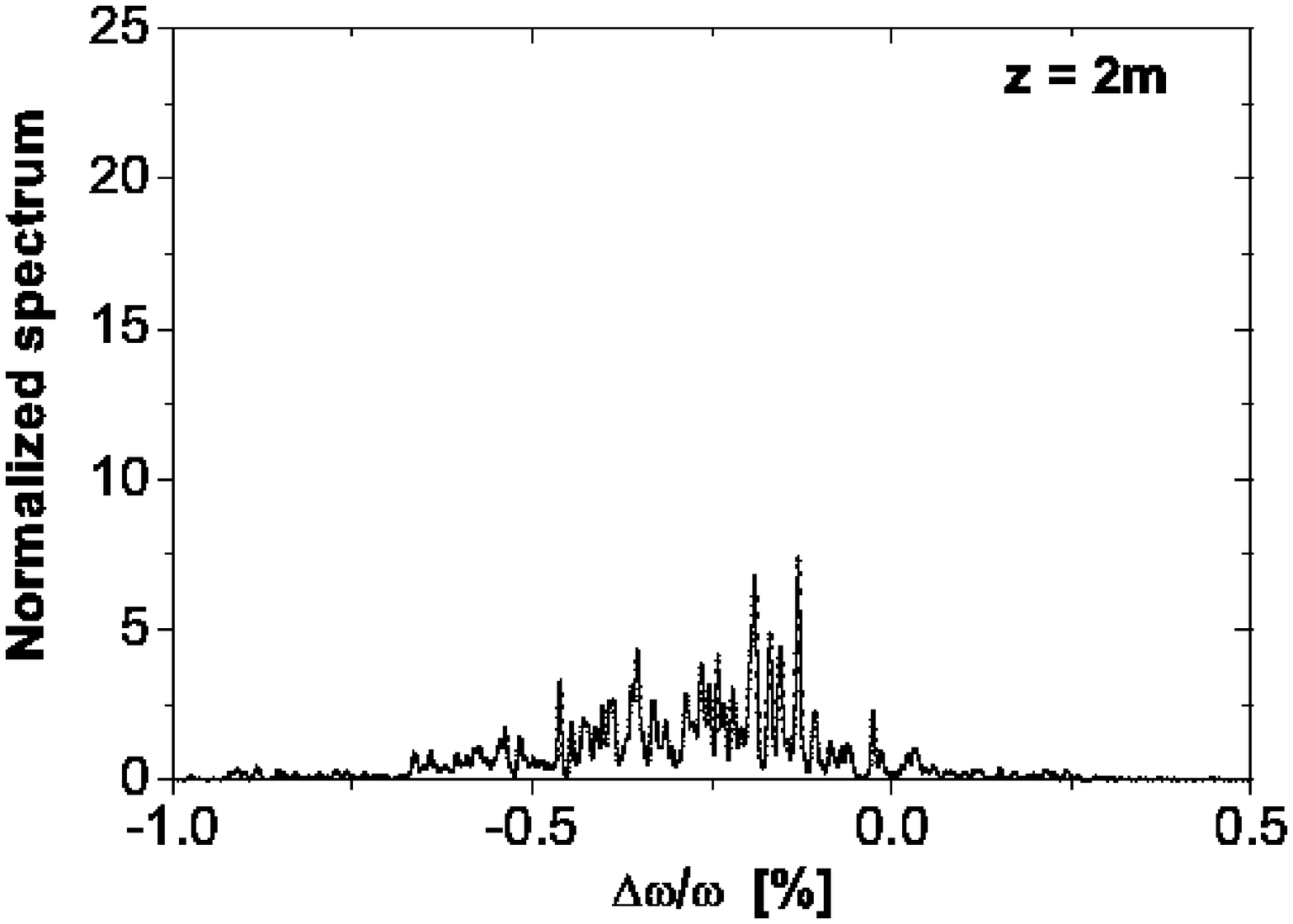,width=0.5\textwidth}

\caption{
Time structure (left plot) and spectral structure (right plot) of the
3~nm radiation pulse for a 2~m long after-burner.
The dashed line shows the bunch profile.
}
\label{fig:p0021006}
\end{figure}

An option of second harmonic generation at TTF FEL using an after-burner 
has been under discussion for several years \cite{faatz}. Here we
present the results of time-dependent simulations of the after-burner
scheme.  The parameters of the electron beam are given in
Table~\ref{tab:beampar}, and the parameters of the undulator and the output
radiation are presented in Table~\ref{tab:after-burner}.

In the after-burner scheme the spent electron beam leaving the main X-ray
undulator passes an undulator tuned to the second harmonic. At the exit
of the main undulator the electron beam has a pronounced amplitude of density
modulation at the second harmonic which serves as input signal for the
second-harmonic undulator (see Fig.~\ref{fig:ab40a2}). When the electron
beam enters the after-burner radiator, it readily starts to produce
radiation. However, the power growth saturates quickly at 2 meters, as it
is seen from Fig.~\ref{fig:ez21}. This is due to the large energy
spread induced in the main undulator (see Fig.~\ref{fig:ab40a2}).
Figure~\ref{fig:p0021006} presents the temporal and spectral structure of
the radiation pulse at the exit of the after-burner. We find that the FWHM
pulse duration is relatively large, about 250~fs. The spectral width is
also large, about 0.4\%, and is driven by the FEL process in the
main undulator. Finally, the level of output radiation power is low,
about 60~MW, i.e. about 3\% of the power of the fundamental frequency of the main undulator.
Remembering that SASE radiation from a planar undulator at saturation
always contains the 3rd harmonic at a per cent level, we conclude that
harmonic generation using an after-burner does not provide extra
opportunities for user applications.

\begin{table}[h]
\caption{\sl
3~nm option with after-burner
}
\medskip

\begin{tabular}{ l l }
\hline
\underline{Undulator}\\
\hspace*{10pt} Type                        & planar \\
\hspace*{10pt} Period                      & 1.95 cm        \\
\hspace*{10pt} Gap                         & 10 mm         \\
\hspace*{10pt} Peak magnetic field         & 0.39 T        \\
\hspace*{10pt} Undulator length            & 2 m         \\
\underline{Coherent radiation} \\
\hspace*{10pt} Wavelength                  & 3 nm       \\
\hspace*{10pt} Energy per pulse            & 15 $\mu $J \\
\hspace*{10pt} Peak power                  & 60 MW       \\
\hspace*{10pt} Bandwidth (FWHM)            & 0.4\%       \\
\hspace*{10pt} Pulse duration (FWHM)       & 250 fs       \\
\hline
\end{tabular}

\label{tab:after-burner}
\end{table}

\end{document}